\documentclass[a4paper,11pt]{article}
\pdfoutput=1 

\usepackage{jcappub} 

\usepackage[T1]{fontenc} 
\usepackage{placeins}
\usepackage[dvipsnames,svgnames,x11names]{xcolor}
\usepackage{array}
\usepackage{graphicx}
\usepackage{booktabs}
\usepackage{slashed} 
\RequirePackage{mathrsfs}
\usepackage{dsfont}
\usepackage{multirow}
\usepackage{hyperref}
\usepackage{cleveref} 
	\crefname{equation}{Eq.}{Eqs.}
	\crefname{figure}{Fig.}{Figs.}
	\crefname{section}{Sec.}{Secs.}
   	\crefname{appendix}{Appendix}{Appendices}

\definecolor{darkgreen}{rgb}{0.2,0.6,0.0}
\definecolor{darkcyan}{rgb}{0.0,0.4,0.8}
\hypersetup{
	colorlinks = true,
	citecolor = darkcyan,
	linkcolor = darkgreen,
	urlcolor = blue
}

\graphicspath{{./plots/}}

\usepackage{floatrow}
\newfloatcommand{capbtabbox}{table}[][\fracBwidth]

\usepackage{ragged2e}
\allowdisplaybreaks

\newcommand{\erf}{\mathop{\mathrm{erf}}}

\newcommand{\kms}{\,\frac{\mathrm{km}}{\mathrm{s}}}
\renewcommand{\epsilon}{\varepsilon}
\renewcommand{\vec}[1]{\boldsymbol{#1}}
\renewcommand{\bar}{\overline}
\renewcommand{\Re}{\mathrm{Re}}
\renewcommand{\Im}{\mathrm{Im}}
\newcolumntype{L}{>{$}l<{$}}
\newcolumntype{C}{>{$}c<{$}}
\newcommand{\N}{\mathcal N}

\newcommand{\vq}{\vec q}
\newcommand{\vv}{\vec v}
\newcommand{\vo}{v_\oplus}
\newcommand{\vvo}{\vec v_\oplus}
\newcommand{\vesc}{v_\text{esc}}
\newcommand{\vx}[1]{v^\perp_{\chi#1}}
\newcommand{\vvx}{\vec v^\perp_\chi}
\newcommand{\vvxSq}{v^{\perp 2}_\chi}

\newcommand{\mmm}[9]{\left[\begin{array}{@{}ccc@{}}#1&#2&#3\\#4&#5&#6\\#7&#8&#9\end{array}\right]}

\title{\boldmath \justifying A general upper bound on the light dark matter scattering rate in materials}

\author[]{Riccardo Catena}
\author[]{and Micha\l~Iglicki}

\affiliation[]{Department of Physics, Chalmers University of Technology, SE-412 96 G\"oteborg, Sweden}

\emailAdd{catena@chalmers.se}
\emailAdd{michal.iglicki@chalmers.se}

\abstract{Combining an effective theory description of spin-1/2 dark matter (DM)-electron interactions in materials with linear response theory provides a powerful framework to model the scattering of DM, including in-medium effects, in detectors used for direct searches. Within this framework, we show that the rate of DM-induced electronic transitions in detector materials admits a theoretical upper bound under general assumptions on the underlying DM-electron coupling. In particular, our theoretical upper bound applies to models where DM couples to the electron density as well as the spin, paramagnetic and Rashba currents in materials, and arises from the Kramers-Kronig relations that constrain the analytic properties of the scattering rate. We evaluate our maximum rate formula numerically for Ar, Xe, Ge, and Si targets and find that Ge and Si detectors are closer to saturate this theoretical upper bound, but still far from saturation
when DM couples to densities or currents which are different from the electron
density. This motivates the exploration of a different class of materials to effectively
probe such coupling forms.} 

\begin{document} 
\maketitle
\flushbottom

\section{Introduction}

The quest for dark matter (DM) is gradually broadening its focus, with a remarkable, global experimental effort directed towards probing models where the DM particle is approximately in the MeV to GeV mass range~\cite{Battaglieri:2017aum,Mitridate:2022tnv,Balan:2024cmq}. The experimental search for DM particles lighter than a GeV is motivated by the lack of discovery of WIMPs at direct detection experiments, and by the possibility of producing a DM candidate in this mass window with the correct cosmological abundance via the freeze-out mechanism (if the Lee-Weinberg bound~\cite{Lee:1977ua} is circumvented by introducing the exchange of a new particle mediator in the relevant number-changing processes).

The search for DM-induced electronic transitions in germanium~\cite{EDELWEISS:2020fxc} and silicon~\cite{SENSEI:2023zdf,DAMIC-M:2023gxo,SuperCDMS:2024yiv}, as well as liquid argon~\cite{DarkSide:2022knj} and xenon~\cite{XENON:2022ltv}, has so far played a major role in the exploration of the MeV to GeV mass window in direct detection experiments. The formalism for describing the scattering of Milky Way, sub-GeV DM particles on the electrons bound to a detector material has been put forward in~\cite{Kopp:2009et,Essig:2011nj,Essig:2015cda}, and subsequently extended to account for in-medium screening in models where DM couples to the electron density in the target~\cite{Hochberg:2021pkt,Knapen:2021run}. More recently, linear response theory has been used to further extend the formalism for DM-electron scattering in materials to effective theories where DM couples to all the electron densities and currents that arise at leading order in the non-relativistic expansion of the DM-electron scattering amplitude~\cite{Catena:2024rym}. Effective theories for DM-electron interactions have been formulated in~\cite{Catena:2019gfa,Catena:2021qsr}, as well as in~\cite{Trickle:2020oki,Krnjaic:2024bdd}.

Focusing on models where fermionic DM couples to the electron density, it has been pointed out that there exists a theoretical upper bound on the rate of DM-induced electronic transitions in a detector~\cite{Lasenby:2021wsc}. This bound arises from the so-called Kramers-Kronig relations, which follow from the analytic properties of the dielectric function, upon which the DM-induced electronic transition rate depends. 
This is an interesting observation, because it provides a systematic framework to identify optimal detector materials to probe the DM coupling to the electron density.

In this work, we show that a theoretical upper bound on the rate of spin-1/2 DM-induced electronic transitions exists not only in models where DM couples to the electron density, but also in models where it couples to the paramagnetic current, the spin current, the scalar product of spin and paramagnetic current, and the Rashba spin-orbit current. This extends the results of~\cite{Lasenby:2021wsc} to the leading currents and densities arising from the non-relativistic expansion of the DM-electron scattering amplitude, or, equivalently, from the non-relativistic expansion of the Dirac Hamiltonian~\cite{Catena:2024rym}.~Analogously to~\cite{Lasenby:2021wsc}, the existence of the upper bound we find here follows from an application of the Kramers-Kronig relations. However, we do not apply the Kramers-Kronig relations to the dielectric function only, but also to the generalized susceptibilities that describe the linear response of materials to the leading, non-relativistic DM-electron couplings. For each given DM coupling, the ratio between the actual DM-induced electronic transition rate and our theoretical upper bound only depends on the assumed detector material and DM mass. For all currents and densities listed above, we evaluate this ratio focusing on Si, Ge, Ar, and Xe as detector materials. In general, we find that Si and Ge are closer to saturate our theoretical upper bound, but still far from saturation in all models where DM couples to densities or currents which are different from the electron number density. This points towards the need for a different class of materials to effectively probe such coupling forms. 

Our work is organized as follows. In \cref{sec:kk} we review the use of the Kramers-Kronig relations in the context of DM direct detection, and show why they imply a theoretical upper bound on the rate of DM-induced electronic transitions in models where DM couples to the electron density in the detector. In \cref{sec:gensus}, we review the application of linear response theory to DM direct detection, and show that, for general DM-material couplings, the rate of DM-induced electronic transitions can be expressed in terms of generalized susceptibilities which, under general conditions, also obey the Kramers-Kronig relations. In \cref{sec:analytic_res}, we use our generalized susceptibility formalism to obtain a theoretical upper bound on the rate of DM-induced electronic transitions in models where DM couples to the leading, non-relativistic electron densities and currents. In \cref{sec:analytic_models}, we provide expressions for the interaction rate calculated in some specific effective models of DM-electron interactions. We evaluate our theoretical upper bound in \cref{sec:numerical}, and conclude in \cref{sec:conclusions}. Additional results and useful identities, as well as a short discussion of some assumptions, are collected in the \cref{app:vDM,app:intdv,app:F,app:assumptions}.

\section{Electronic transitions induced by a DM-electron density coupling}
\label{sec:kk}

As already mentioned in the introduction, throughout the whole paper we assume the electrons to be non-relativistic, so that their kinetic energies are much smaller than $m_e$. Nevertheless, some electrons in the material, namely, those at the innermost atomic shells, may be predominantly relativistic, as their expected kinetic energies are comparable to $m_e$. Those electrons, however, are tightly bounded to the nucleus and, therefore, difficult to excite. Therefore, they do not significantly contribute to the total excitation rate. The valence electrons, whose excitations would be the main source of the DM-induced signal, are not subject to any significant relativistic corrections \cite{Tatewaki2017}, so the non-relativistic treatment is justified.

\subsection{Fermi's golden rule}

The rate of transitions from an electronic state $|i\rangle$ to an electronic state $|f\rangle$, induced by an incoming DM particle of momentum $\vec p$ and a spin configuration $s$ scattering to a final state momentum between $\vec p'$ and $\vec p' + {\rm d} \vec p'$ and with a spin configuration $s'$, is given by Fermi's golden rule, 
\begin{align}
{\rm d}\Gamma^{ss'}_{i\rightarrow f}(\vec p) = (2 \pi) \delta(E_f+E_{\vec p'} - E_i-E_{\vec p}) \left| \langle f; \vec p',s' |\,\widehat{V}\,|\vec p, s; i \rangle \right|^2 \, \frac{V {\rm d} \vec p' }{(2\pi)^3} \,,
\label{eq:fermi}
\end{align}
where $\widehat{V}$ is the DM-electron interaction potential, $E_i$ ($E_f$) the energy of the initial (final) electronic state, and $E_{\vec p}$ ($E_{\vec p'}$) the energy of the initial (final) DM particle. Here, single particle states are normalized to one, while $V\equiv \int {\rm d}\vec x$ is a normalization volume such that the number of DM particle states with momenta between $\vec p'$ and $\vec p' + {\rm d} \vec p'$ in a volume $V$ is $V {\rm d} \vec p'/(2\pi)^3$.

In position space, the matrix element of the interaction potential $\widehat{V}$ can be written as follows,
\begin{align}
\langle f;\vec p',s'  | \widehat{V} | \vec p,s; i \rangle =& \frac{1}{V} \int{\rm d}\vec r_e   \int{\rm d}\vec r_\chi
\,\psi_f^{*}(\vec r_e) \,e^{- i \vec p'\cdot \vec r_\chi} \, \xi_\chi^{s'\dagger} \widehat{V}_{x}(\vec r_e,\vec r_\chi,\dots)  \xi_\chi^s \, e^{i \vec p\cdot \vec r_\chi}  \,  \psi_i(\vec r_e)\,,
\label{eq:Vmatrix}
\end{align}
where we have assumed spin-1/2 DM and $\xi_\chi^s$ ($\xi_\chi^s$) is a two-component spinor describing the spin configuration of the initial (final) DM particle (i.e., up or down for spin-1/2 DM), $\psi_i$ ($\psi_f$) is the initial (final) electronic wave function, while $\widehat{V}_{x}$ denotes the position space interaction potential. The dots in \cref{eq:Vmatrix} denote gradient and spin operators which $\widehat{V}_{x}$ can in general depend on. In \cref{eq:Vmatrix}, we integrate over the electron and DM particle position vectors, $\vec r_e$ and $\vec r_\chi$.

By assuming $\widehat{V}_{x}(\vec r_e,\vec r_\chi,\dots) = \mathds{1}_e\mathds{1}_\chi\widehat{V}_{x}(\vec r_e-\vec r_\chi)$ (which applies to the case of velocity- and spin-independent interactions), where $\mathds{1}_e$ ($\mathds{1}_\chi$) is the $2\times2$ identity in the electron (DM particle) spin space, the matrix element in \cref{eq:Vmatrix} can now be rewritten as follows,
\begin{align}
\label{eq:VV}
\begin{aligned}
\langle f;\vec p',s'  | \widehat{V} | \vec p,s; i \rangle
&= \int{\rm d}\vec r_e \,\psi_f^{*}(\vec r_e) V^{ss'}_{\rm eff}(\vec r_e, \vec q) \, \psi_i(\vec r_e) \\
&=\langle f | \,V^{ss'}_{\rm eff} \, | i \rangle\,,
\end{aligned}
\end{align}
where
\begin{align}
V^{ss'}_{\rm eff}(\vec r_e,\vec q) &\equiv \frac{1}{V}  
e^{i \vec q\cdot \vec r_e}  \, \mathds{1}_e \,\delta^{ss'} \widetilde{V}_{x}(\vec q)\,,
\label{eq:Veff}
\end{align}
is the effective transition potential,
\begin{align}
 \widetilde{V}_{x}(\vec q) \equiv \int {\rm d} \left(\vec r_e-\vec r_\chi \right) \, e^{-i \vec q \cdot (\vec r_e-\vec r_\chi)} \, \widehat{V}_{x}(\vec r_e-\vec r_\chi) \,,
\end{align}
and $\vec q = \vec p - \vec p'$ is the momentum transfer. Recalling that the number density of an electron at $\vec r_e$ and its Fourier transform at $\vec q$ are given by
\begin{align}
n_0(\vec r) &= \delta^{(3)}(\vec r - \vec r_e)\,,&
\widetilde{n}_0(\vec q) &= e^{-i \vec q \cdot \vec r_e} \,,
\end{align}
we find that
\begin{align}
V^{ss'}_{\rm eff}(\vec r_e,\vec q) &\equiv \frac{1}{V}  \,
\widetilde{n}_0(-\vec q) \, \mathds{1}_e \, \delta^{ss'} \widetilde{V}_{x}(\vec q) \,.
\label{eq:Veff_n}
\end{align}
Independently of the specific form of $\widetilde{V}_{x}(\vec q)$, whenever $\widehat{V}_{x}(\vec r_e,\vec r_\chi,\dots) = \mathds{1}_e \mathds{1}_\chi \widehat{V}_{x}(\vec r_e-\vec r_\chi)$ the transition potential $V^{ss'}_{\rm eff}(\vec r_e,\vec q)$ depends on the properties of the electrons bound to the target material through the electron density solely. One can summarize this property of the underlying DM-electron interaction by saying that the DM couples to the electron density in the material.

As an example, let us consider the following amplitude for DM scattering by free electrons, 
\begin{align}
\mathscr{M}= c_1 \, \delta^{ss'} \delta^{rr'} \,,
\label{eq:M}
\end{align}
where $c_1$ is a dimensionless constant, while  $r$ ($r'$) labels the initial (final) state electron spin. In the Born approximation, the relation between $\mathscr{M}$ and the associated scattering potential is
\begin{align}
\langle \vec k',r';\vec p',s'  | \widehat{V} | \vec p,s; \vec k,r  \rangle &= - (2\pi)^3 \delta^{(3)}(\vec p'+\vec k' - \vec k - \vec p)\, \frac{\mathscr{M}}{4 m_e m_\chi V^2} \,,
\label{eq:B}
\end{align}
where $\vec k$ ($\vec k'$) is the initial (final) free electron momentum. By inserting \cref{eq:Vmatrix} with free electron wave functions $\psi_i(\vec r_e)= \xi_e^r e^{i \vec k\cdot \vec r_e}/\sqrt{V}$ and $\psi_f(\vec r_e)= \xi_e^{r'} e^{i \vec k'\cdot \vec r_e}/\sqrt{V}$ into \cref{eq:B}, we obtain
\begin{align}
\widehat{V}_{x}(\vec r_e-\vec r_\chi) = - \frac{c_1}{4 m_e m_\chi} \, \delta^{(3)}(\vec r_e-\vec r_\chi) \,,
\end{align}
and, using \cref{eq:Veff_n},
\begin{align}
V^{ss'}_{\rm eff}(\vec r_e,\vec q) = - \frac{c_1}{4 m_e m_\chi V} \, \delta^{ss'} \, \mathds{1}_e \, \widetilde{n}_0(-\vec q)\,.
\end{align}

\subsection{Dielectric function formalism}
By inserting \cref{eq:VV}, with $V_{\rm eff}^{ss'}(\vec r_e,\vec q)$ given in \cref{eq:Veff_n}, into \cref{eq:fermi}, we find the rate formula
\begin{align}
{\rm d}\Gamma^{ss'}_{i\rightarrow f}(\vec p) = \frac{(2 \pi)}{V}\, \delta(E_f+E_{\vec p'} - E_i-E_{\vec p}) \,\delta^{ss'}\,| \widetilde{V}_{x}(\vec q) |^2 \,  \left| \langle f |\,\widetilde{n}_0(-\vec q)\,| i \rangle \right|^2 \, \frac{{\rm d} \vec p' }{(2\pi)^3} \,.
\label{eq:fermi2}
\end{align}
Summing (averaging) over the final (initial) electronic configurations and DM particle spins, and averaging over the initial DM particle velocities $\vec v= \vec p/m_\chi$ (defined in the laboratory frame), we can now calculate the total electronic transition rate induced by a galactic DM particle whose initial velocity in the laboratory frame is distributed according to the probability density $f(\vec v)$, and whose final momentum lies between $\vec p'$ and $\vec p'+{\rm d} \vec p'$. In addition, by integrating over the final DM momenta $\vec p'$ or, equivalently, over the momentum transfer 
$\vec q$, we find
\begin{align}
\Gamma = \frac{1}{2}\sum_{ss'} \, \sum_{if} \, \frac{e^{-\beta E_i}}{Z}  \, \int {\rm d}\vec v \, f(\vec v) \, \int {\rm d}\vec q\, \frac{{\rm d}\Gamma^{ss'}_{i\rightarrow f}(\vec p)}{{\rm d} \vec q}  \,,
\label{eq:fermi3}
\end{align}
where  $Z$  is the partition function, $\beta=1/T$, and $T$ denotes the temperature (in DM direct detection applications, $T\rightarrow 0$; see \cref{app:temperature} for a short discussion). Introducing $K_{n_0 n_0}(\vec q,\omega)$, the density-density correlation function,
\begin{align}
K_{n_0 n_0}(\vec q,\omega) &=\frac{2\pi}{V} \sum_{if} \frac{e^{-\beta E_i}}{Z} \,
\langle f|  \widetilde{n}_0(-\vec q) |i\rangle\, \langle i| \widetilde{n}_0(\vec q) |f\rangle \,\delta(E_f-E_i-\omega)\,,
\label{eq:corr}
\end{align}
we can rewrite $\Gamma$ as
\begin{align}
\Gamma =  \int \frac{{\rm d} \vec q }{(2\pi)^3} \, | \widetilde{V}_x(\vec q) |^2  \int {\rm d}\vec v \, f(\vec v) \,
K_{n_0 n_0}(\vec q,\omega_{\vec v, \vec q} )  \,,
\end{align}
where
\begin{align}
 \omega_{\vec v, \vec q} &= E_{\vec p} - E_{\vec p - \vec q} = \vec q \cdot \vec v - \frac{q^2}{2 m_\chi} \,.
\label{eq:omega}
\end{align}
Finally, by using~\cite{Catena:2024rym}%
\footnote{The local field effects, discussed in \cite{Catena:2024rym}, are neglected here.}
\begin{align}
K_{n_0 n_0}(\vec q,\omega) = \frac{2}{1-e^{-\beta \omega}} \, \Im\left[ - \epsilon_r(\vec q,\omega)^{-1} \right] \, U(q)^{-1}
\,,
\label{eq:Knn}
\end{align}
where $U(q)=4\pi \alpha/q^2$ is the Fourier transform of the Coulomb potential and $\alpha$ the fine structure constant, we obtain
\begin{align}
\Gamma &= \int \frac{{\rm d} \vec q }{(2\pi)^3}  \, | \widetilde{V}_x(\vec q) |^2  \int {\rm d}\vec v \, f(\vec v) \,
 \frac{2}{1-e^{-\beta \omega_{\vec v, \vec q}}} \, {\rm Im} \left[ - \epsilon_r(\vec q,\omega_{\vec v, \vec q})^{-1} \right] \,  U(q)^{-1}
 \,,
 \label{eq:Reps}
\end{align}
which gives a formula for the DM-induced electronic transition rate in terms of the dielectric function $\epsilon_r(\vec q,\omega)$ (defined as in \cite{Solyom2011-rp}). \Cref{eq:Reps} applies to models where DM couples to the electron density in materials.

\subsection{Kramers-Kronig relations}
By combining the dielectric function formalism reviewed above with the Kramers-Kronig relations, one can derive a theoretical upper bound on $\Gamma$ in models where DM couples to the electron density in materials~\cite{Lasenby:2021wsc}. To show this, let us first introduce the Kramers-Kronig relations.

Let function $g:\mathbb{C} \rightarrow \mathbb{C}$ be analytic in the upper half plane,\footnote{This is satisfied if the Fourier transform of $g$ is proportional to the step function. Physically, it means that if $z$ denotes frequency, the Fourier transform of $g$ is causal as a function of time.} and $g(z) \rightarrow 0$ for $|z|\rightarrow \infty$. Then, $g$ must satisfy the following Kramers-Kronig relations \cite{Solyom2011-rp}:
\begin{subequations}
\begin{align}
{\rm Re} \, g(z_0)& = \frac{1}{\pi} \mathcal{P} \int_{-\infty}^{+\infty}\frac{ {\rm d}z}{z-z_0}\, {\rm Im} \, g(z) \,,\\
{\rm Im} \, g(z_0)& = -\frac{1}{\pi} \mathcal{P} \int_{-\infty}^{+\infty}\frac{ {\rm d}z}{z-z_0}\, {\rm Re} \, g(z) \,,
\end{align}
\end{subequations}
where $\mathcal{P}$ denotes the principal value. In particular, for $z_0=0$, we obtain
\begin{subequations}
\label{eq:g} 
\begin{align}
{\rm Re} \, g(0)& = \frac{1}{\pi} \mathcal{P} \int_{-\infty}^{+\infty}\frac{ {\rm d}z}{z}\, {\rm Im} \,g(z)\,, \label{eq:g1}\\
{\rm Im} \, g(0)& = - \frac{1}{\pi} \mathcal{P} \int_{-\infty}^{+\infty} \frac{{\rm d}z}{z}\, {\rm Re} \,g(z) \,.\label{eq:g2}
\end{align}
\end{subequations}
If, in addition, $g$ satisfies a third property---namely, that its Fourier transform is real%
 \footnote{For example, if $z$ is frequency, then this third condition requires the function $g$ to be real in the time domain.}%
 ---its real part ${\rm Re}\, g(z)$ is even and its imaginary part ${\rm Im}\, g(z)$ is odd as a function of $z\in\mathbb{R}$. Then, \cref{eq:g1} becomes 
\begin{align}
 \int_{0}^{+\infty} \frac{{\rm d}z}{z}\,{\rm Im} \,g(z) = \frac{\pi}{2} g(0) \,. 
\end{align}
Since $1- \epsilon_r(\vec q,\omega)^{-1}$ as a function of $\omega$ obeys the above three conditions for each $\vec q$, one can write
\begin{align}
 \int_{0}^{+\infty} \frac{{\rm d}\omega}{\omega}\,{\rm Im} \left[ 1- \epsilon_r(\vec q,\omega)^{-1}\right] = \frac{\pi}{2} \left[ 1- \epsilon_r(\vec q,0)^{-1} \right] \,. 
 \label{eq:KKeps}
\end{align}
This identity implies a theoretical upper bound on the rate of DM-induced electronic transitions $\Gamma$~\cite{Lasenby:2021wsc}. Indeed, in \cref{eq:Reps}, the change of integration variables introduced in \cref{app:intdv} leads to
\begin{align}
\Gamma = \int_0^\infty {\mathrm d}q \,\frac{q^2 \, U(q)^{-1}}{2\pi^2} \, | \widetilde{V}_x(q) |^2 \,\int_0^\infty {\mathrm d}\omega\,\rho^{(0)}(\omega,q)\, 
 \frac{2}{1-e^{-\beta \omega}} \, {\rm Im} \left[ - \epsilon_r(q,\omega)^{-1}\right]
\,,
\label{eq:G}
\end{align}
where
\begin{align}
\rho^{(0)}(\omega,q)
= \frac{\pi}{q}\int_{v_q}^\infty {\mathrm d}v \,v \int_{-1}^1 {\mathrm d}\cos\alpha\,f(\vec v)\,.
\end{align}
Here, we assume
\begin{align}\label{eq:f}
\begin{aligned}
f(\vec v)
&=f(v,\cos\alpha)\\
&=\N\,\exp\left[-\frac{(\vv + \vvo)^2}{v_0^2}\right]\,\theta(\vesc-|\vv+\vvo|)\,,
\end{aligned}
\end{align}
where $\vec v \cdot \vvo = v \vo \cos\alpha$, $\vvo$ is Earth's velocity relative to the galactic centre, $\vesc$ is the galactic escape velocity, $v_0$ is the most probable DM speed (in the galactic reference frame, in which the mean DM velocity is zero), while $\N$ is a normalization constant. The assumed values of $\vo\equiv|\vvo|$, $\vesc$, $v_0$, and $\N$ are listed in \cref{app:vDM}. We also introduce the minimum DM speed required to deposit an energy $\omega$,
\begin{align}
\label{eq:vq}
\begin{aligned}
v_q
&\equiv \vec v\cdot\frac{\vec q}{q}\\
&=  \frac{\omega}{q} + \frac{q}{2\,m_\chi}\,,
\end{aligned}
\end{align}
and assume an isotropic target (equivalently: average over detector's orientations, see \cref{app:isotropy} for a short discussion), so that we can ignore the dependence of $\epsilon_r$ on the direction of $\vec q$,%
\footnote{If DM couples to the electron density solely, invariance under three-dimensional rotations implies ${\widetilde{V}_x(\vec q) = \widetilde{V}_x(q)}$.}
\begin{align}
\label{eq:epsilon}
\epsilon_r(\vec q,\omega) \simeq \epsilon_r(q,\omega) \,.
\end{align}
\Cref{eq:KKeps,eq:G} can now be combined to obtain a theoretical upper bound, $\Gamma_{\rm opt}$, on~$\Gamma$~\cite{Lasenby:2021wsc}:
\begin{align}
\Gamma \le \Gamma_{\rm opt} =
\int_0^\infty {\mathrm d}q   \,
 \frac{q^2\, U(q)^{-1}}{2 \pi} \, | \widetilde{V}_x(q) |^2 \, \max_{\omega} \left[ \omega \rho^{(0)}(\omega,q) \right] \,,
\end{align}
where we took into account that for $T\rightarrow 0$, $1-e^{-\beta \omega} \rightarrow 1$, and, following~\cite{Lasenby:2021wsc}, assumed $1 - \epsilon_r(q,0)^{-1} \le 1$ and $\Im[-\epsilon_r(q,\omega)^{-1}]\ge 0$.

\section{Electronic transitions induced by general DM-electron couplings}
\label{sec:gensus}

\subsection{Fermi's golden rule}

\noindent In general, DM can couple to electron densities and currents in materials that are different from the electron number density $n_0$. Specifically, the most general form for the transition potential $V^{ss'}_{\rm eff}$ at leading order in the electron velocity and in the momentum transfer is given by~\cite{Catena:2024rym}
\begin{align}
\begin{aligned}
V^{ss'}_{\rm eff}(\vec r_e,\vec q) = -\frac{1}{4m_e m_\chi V}\,\Big\{\mkern20mu
&F_{n_0}^{ss'}e^{i\vec q\cdot \vec r_e}\,\mathds{1}_{e}\\
+\,&F_{n_A}^{ss'} \frac{i}{2m_e}\left[ \overleftarrow{\nabla}_{\vec r_e}\cdot \boldsymbol{\sigma}_e\, e^{i\vec q\cdot{\vec r_e}} -  e^{i\vec q\cdot{\vec r_e}} \,\boldsymbol{\sigma}_e\cdot\overrightarrow{\nabla}_{\vec r_e} \right] \\
+\,&\vec F_{\vec j_5}^{ss'}\cdot \boldsymbol{\sigma}_e\,e^{i\vec q\cdot \vec r_e} \\
+\,&\vec F_{\vec j_M}^{ss'}\cdot  \frac{i}{2m_e}\left[ \overleftarrow{\nabla}_{\vec r_e} e^{i\vec q\cdot{\vec r_e}} -  e^{i\vec q\cdot{\vec r_e}}\overrightarrow{\nabla}_{\vec r_e} \right] \mathds{1}_{e}\\
+\,&\vec F_{\vec j_E}^{ss'}\cdot  \frac{1}{2m_e}\left[ \overleftarrow{\nabla}_{\vec r_e}\times\boldsymbol{\sigma}_e\, e^{i\vec q\cdot{\vec r_e}} +  e^{i\vec q\cdot{\vec r_e}} \,\boldsymbol{\sigma}_e\times\overrightarrow{\nabla}_{\vec r_e} \right]
\Big\} \,,
\label{eq:Vefftot}
\end{aligned}
\end{align}
where an arrow on the gradient operator indicates whether it acts on the initial ($\overrightarrow{\nabla}_{\vec r_e}$) or final  ($\overleftarrow{\nabla}_{\vec r_e}$) electron wave function, while the pre-factors $F_{n_0}^{ss'}$, $F_{n_A}^{ss'}$,  $\vec F_{\vec j_5}^{ss'}$, $\vec F_{\vec j_M}^{ss'}$, and $\vec F_{\vec j_E}^{ss'}$ are model dependent. These pre-factors are listed in \cref{app:F} for the effective theory of DM-electron interactions of~\cite{Catena:2019gfa,Catena:2021qsr} and in \cref{sec:anapole,sec:magdip,sec:eldip} for specific models where DM has as an anapole, electric dipole, or magnetic dipole moment.

The first line in \cref{eq:Vefftot} corresponds to the already discussed case of models where DM couples to the electron number density in materials. The third and fourth lines describe a coupling between DM and the electron spin or paramagnetic current, respectively. The second line can be identified with the spin-paramagnetic current coupling, and the last line with the Rashba term arising at second order in the $1/c$ expansion of the Dirac Hamiltonian~\cite{Catena:2024rym}.

Inserting \cref{eq:Vefftot} into \cref{eq:VV}, and then the latter into Fermi's golden rule, \cref{eq:fermi}, we obtain the following expression for the rate of DM-induced electronic transitions in materials,
\begin{align}
\Gamma &=  \frac{1}{16 m_e^2 m_\chi^2} \,\sum_{a b} \int \frac{{\rm d}\vec q}{(2 \pi)^3} \int {\rm d}\vec v \, f(\vec v)  \, \mathcal{F}_{ab}(\vec q,\vec v)
K_{b^\dagger a }(\vec q,\omega_{\vec v,\vec q}) \,,
\label{eq:rate}
\end{align}
where
\begin{align}
\mathcal{F}_{ab}(\vec q,\vec v) = \frac{1}{2}\sum_{s s'} F^{ss'}_a(\vec q,\vec v)\, F_b^{ss'*}(\vec q,\vec v) \,,
\label{eq:F2}
\end{align}
and $K_{b^\dagger a }(\vec q,\omega)$ is the correlation function,
\begin{align}
K_{b^\dagger a }(\vec q,\omega) = \frac{2\pi}{V} \sum_{if} \frac{e^{-\beta E_i}}{Z}\,
\langle f|  \widetilde{a}(-\vec q) |i\rangle\, \langle i| \widetilde{b}^\dagger(\vec q) |f\rangle\, \delta(E_f-E_i-\omega)\,,
\label{eq:corrgen}
\end{align}
while $a$ and $b$ could be any of the densities, or components of currents, contributing to \cref{eq:Vefftot}, namely
\begin{align}
(\widetilde{n}_0, \,\, \widetilde{n}_A, \,\, \widetilde{j}_{5_1},  \,\, \widetilde{j}_{5_2}, \,\, \widetilde{j}_{5_3}, \,\, \widetilde{j}_{M_1}, \,\, \dots. \,\, \widetilde{j}_{E_1}, \,\, \dots)\,,
\label{eq:notation}
\end{align}
where
\begin{subequations}
\begin{align}
\widetilde{n}_0(\vec q) &\equiv e^{-i \vec q \cdot \vec r_e} \,,\\
\widetilde{n}_A(\vec q) &\equiv  \frac{i}{2m_e}\left[ \overleftarrow{\nabla}_{\vec r_e}\cdot \boldsymbol{\sigma}_e\, e^{-i\vec q\cdot{\vec r_e}} -  e^{-i\vec q\cdot{\vec r_e}} \,\boldsymbol{\sigma}_e\cdot\overrightarrow{\nabla}_{\vec r_e} \right]  \,,\\
\widetilde{\vec j}_5(\vec q) &\equiv \boldsymbol{\sigma}_e\,e^{-i\vec q\cdot \vec r_e} \,,\\
\widetilde{\vec j}_M(\vec q) &\equiv \frac{i}{2m_e}\left[ \overleftarrow{\nabla}_{\vec r_e} e^{-i\vec q\cdot{\vec r_e}} -  e^{-i\vec q\cdot{\vec r_e}}\overrightarrow{\nabla}_{\vec r_e} \right] \,,\\
\widetilde{\vec j}_E(\vec q) &\equiv \frac{1}{2m_e}\left[ \overleftarrow{\nabla}_{\vec r_e}\times\boldsymbol{\sigma}_e\, e^{-i\vec q\cdot{\vec r_e}} + e^{-i\vec q\cdot{\vec r_e}} \,\boldsymbol{\sigma}_e\times\overrightarrow{\nabla}_{\vec r_e} \right] \,.
\end{align}
\end{subequations}
We denote by $\boldsymbol{\sigma}_e=(\sigma_1, \sigma_2, \sigma_3)$ the three-dimensional vector whose components are the Pauli matrices.
For increased readability, in the remaining part of this work, we omit the tildes (e.g., we write $n_0$ instead of $\widetilde n_0$). In other words, all the densities and currents are hereafter meant to be expressed in momentum space.

\subsection{Generalized susceptibility formalism}
As the density-density correlation function $K_{n_0n_0}$ in \cref{eq:Knn} is related to the dielectric function, so is the correlation function $K_{b^\dagger a}$ related to generalized susceptibilities describing the linear response of materials to a perturbation induced by general DM-electron couplings~\cite{Catena:2024rym}.

Indeed, in the interaction picture, $V_{\rm eff}^{ss'}$ can be written as a time dependent perturbation~\cite{Catena:2024rym}:
\begin{align}
V_{\rm eff}^{ss'}(t) = - \sum_{a}\int {\rm d}\vec r \, B_a(\vec r) \, S_a^{ss'}(\vec r,t) \,,
\label{eq:Vt}
\end{align}
with
\begin{align}
B_a(\vec r)  = \int \frac{{\rm d}\vec q'}{(2\pi)^3} \, e^{i \vec q' \cdot \vec r} \, a(\vec q') \,,
\end{align}
and 
\begin{align}
S_a^{ss'}(\vec r,t) = \frac{1}{4 m_e m_\chi V} \,F_a^{ss'} \, e^{i \vec q \cdot \vec r} \,e^{-i\omega_{\vec v, \vec q} t} \,.
\end{align}
Notice that at $t=0$ \cref{eq:Vt,eq:Vefftot} coincide. In linear response theory, the fluctuation, $\langle \Delta a(\vec r,t)\rangle$, induced on the generic electron density or component $a$ by the potential $V_{\rm eff}^{ss'}(t)$  
is given by 
\begin{align}
\langle \Delta a(\vec r,t)\rangle &= \sum_{b}\int_{-\infty}^{t} {\rm d} t' \int {\rm d}\vec r'\, \chi_{a b}(\vec r-\vec r',t-t') \,S_b^{ss'}(\vec r',t') \,,
\label{eq:dcs}
\end{align}
where
\begin{align}
\chi_{ab}(\vec r-\vec r',t-t') \equiv i\theta(t-t') \Big \langle \left[ a(\vec r,t),b(\vec r',t') \right] \Big \rangle 
\label{eq:chiret}
\end{align}
is the generalized susceptibility associated with $a$ and $b$. The double Fourier transform of $\chi_{ab}$ obeys~\cite{Catena:2024rym}
\begin{align}
\chi_{ab} (\vec q,\omega)-\chi^*_{b^\dagger a^\dagger} (\vec q,\omega) = i K_{ab} (\vec q,\omega)\left( 1- e^{-\beta \omega} \right) \,.
\label{eq:chi1}
\end{align}
In the $T\rightarrow 0$ limit (see a short discussion in \cref{app:temperature}), when $1- e^{-\beta \omega} \rightarrow 1$, we can use \cref{eq:chi1} to rewrite the transition rate $\Gamma$, given by  \cref{eq:rate}, as
\begin{align}
\Gamma &=  \frac{-i}{16 m_e^2 m_\chi^2} \,\sum_{a b} \int \frac{{\rm d}\vec q}{(2 \pi)^3} \int {\rm d}\vec v \, f(\vec v)  \, \mathcal{F}_{ab}(\vec q,\vec v)
\left( \chi_{a^\dagger b} -\chi^*_{b^\dagger a} \right) (\vec q,\omega_{\vec v, \vec q})  \,,
\label{eq:rategen}
\end{align}
with the DM velocity distribution function $f$ defined by \cref{eq:f} and $\mathcal{F}_{ab}$ given by \cref{eq:F2}.

\subsection{Kramers-Kronig relations}
In the time domain, the generalized susceptibilities $\chi_{a^\dagger b}(\vec q,t-t')$ are causal, as one can see from their definition, \cref{eq:chiret}. In particular, this implies $\chi_{a^\dagger b}(\vec q,t-t')=0$ for $t-t'<0$. Consequently, in the (complex) frequency domain $\chi_{a^\dagger b}(\vec q,\omega)$ is analytic in the upper-half plane, that is, for ${\rm Im}\,\omega \ge 0$. Apart of that, $\chi_{a^\dagger b}(\vec q,\omega) \rightarrow 0$ for $|\omega| \rightarrow \infty$. Moreover, ${\chi_{a^\dagger b}(\vec q,t-t')}$ is real in the time domain, which implies that, as a function of $\omega\in\mathbb{R}$, ${\rm Re}\,\chi_{a^\dagger b}(\vec q,\omega)$ is even while ${\rm Im}\,\chi_{a^\dagger b}(\vec q,\omega)$ is odd.

Since $\chi_{a^\dagger b}(\vec q,\omega)$ meets the above three conditions for every $\vec q$, it also obeys \cite{Solyom2011-rp}
\begin{subequations}\label{eq:KK-general}
\begin{align}
{\rm Re} \, \chi_{a^\dag b}(\vec q,\omega)& = \frac{1}{\pi} \mathcal{P} \int_{-\infty}^{+\infty}\frac{ {\rm d}\omega'}{\omega'-\omega}\, {\rm Im} \, \chi_{a^\dag b}(\vec q,\omega') \,,\\
{\rm Im} \, \chi_{a^\dag b}(\vec q,\omega)& = -\frac{1}{\pi} \mathcal{P} \int_{-\infty}^{+\infty}\frac{ {\rm d}\omega'}{\omega'-\omega}\, {\rm Re} \, \chi_{a^\dag b}(\vec q,\omega') \,,
\end{align}
\end{subequations}
as well as
\begin{align}\label{eq:KKgen2}
\int_{0}^{+\infty} \frac{{\rm d}\omega}{\omega}\,{\rm Im} \,\chi_{a^\dagger b}(\vec q,\omega) = \frac{\pi}{2} \chi_{a^\dagger b}(\vec q,0) \,. 
\end{align}
The identities in \cref{eq:KK-general} are the Kramers-Kronig relations for the generalized susceptibilities $\chi_{a^\dagger b}(\vec q,\omega)$, while \cref{eq:KKgen2} extends the relation in \cref{eq:KKeps} to the case of general DM couplings.

\section{Upper bound on the transition rate for general DM-electron couplings}
In this section, we first analytically derive a general upper bound on the DM-induced electronic transition rate (\cref{sec:analytic_res}) and provide expressions for the interaction rate in specific effective models of DM (\cref{sec:analytic_models}).
We then numerically evaluate the transition rate to associated upper bound ratio focusing on germanium, silicon, argon and xenon detectors, and on models where DM exhibits an anapole, magnetic dipole or electric dipole moment (\cref{sec:numerical}). This numerical analysis will allow us to assess which material is closest to saturate our general upper bound on $\Gamma$.

\subsection{Derivation from the Kramers-Kronig relations}
\label{sec:analytic_res}
We start from an analytic derivation of a general upper bound on $\Gamma$ based on the generalized Kramers-Kronig relations that apply to material responses beyond the familiar dielectric function.

Using \cref{eq:KKgen2}, we now show that the sum of \cref{eq:rategen} is bounded from above. Let us recall \cref{eq:rategen}:
\begin{align}
\label{eq:rategenrecalled}
\Gamma &=  \frac{-i}{16 m_e^2 m_\chi^2} \,\sum_{a b} \int \frac{{\rm d}\vec q}{(2 \pi)^3} \int {\rm d}\vec v \, f(\vec v)  \, \mathcal{F}_{ab}(\vec q,\vec v)
\left(\chi_{a^\dag b} - \chi^*_{b^\dag a}\right) (\vec q,\omega_{\vec v, \vec q})  \,.
\end{align}
First, let us notice that the sum $\sum_{ab}\mathcal{F}_{ab}(\chi_{a^\dag b}-\chi_{b^\dag a}^*)$ consists of contributions of the following types:
\begin{subequations}
\label{eq:Fnn-nj-jj}
\begin{align}
&\mathcal{F}_{nn'}\,(\chi_{n^\dag n'}-\chi_{n^{\prime\dag} n}^*)\,,&
&\text{where }n,n'=n_0,n_A\,,\label{eq:Fnn}\\
&\sum_k\left[
    \mathcal{F}_{j_kn}\,(\chi_{j_k^\dag n}-\chi_{n^\dag j_k}^*)
    + \mathcal{F}_{nj_k}\,(\chi_{n^\dag j_k}-\chi_{j_k^\dag n}^*)
    \right]\,,&
&\text{where }\begin{cases}n=n_0,n_A\\\vec j = \vec j_5,\vec j_M,\vec j_E\end{cases}\,,\label{eq:Fnj}\\
&\sum_{k\ell}\mathcal{F}_{j_kj'_\ell}\,(\chi_{j_k^\dag j'_\ell}-\chi_{j^{\prime\dag}_\ell j_k}^*)\,,&
&\text{where }\vec j, \vec j' = \vec j_5,\vec j_M,\vec j_E\,.\label{eq:Fjj}
\end{align}
\end{subequations}
Investigating the explicit forms of $\mathcal{F}_{ab}$ coefficients provided in \cref{app:F}, one can realize that $\mathcal{F}_{n_0n_0}$, the only non-vanishing coefficient of the type $\mathcal{F}_{nn'}$ appearing in \cref{eq:Fnn}, is a~scalar quantity depending on $q\equiv |\vec q|$, $v\equiv |\vec v|$, and $v_q$ defined by \cref{eq:vq}. In the contribution of type \labelcref{eq:Fnj}, the relevant%
\footnote{In the explicit form of $\mathcal{F}_{j_{M_k}n_0}$ and $\mathcal{F}_{n_Aj_{5_k}}$, there appear linear terms of the third type, namely, those proportional to $\epsilon_{kij}\,(q_i/q)\,v_j$. However, they vanish in the models considered by us (cf. \cref{sec:anapole,sec:magdip,sec:eldip}).}
coefficients $\mathcal{F}_{j_kn}(\vq,\vec v)$ depend on the components of $\vec q$ and $\vec v$ in the following way:
\begin{align}
\label{eq:Fjn}
\mathcal{F}_{j_kn}(\vq,\vec v) \equiv A_{\vec jn}(q,v,v_q)\,\frac{q_k}{q} + B_{\vec jn}(q,v,v_q)\,v_k\,.
\end{align}
When deriving our results, we assume that the target material is isotropic (see \cref{app:isotropy}). Consequently, when contracting the $F_{j_kn}$ with the corresponding susceptibility, only the component of $\vec v$ parallel to $\vec q$ is relevant. Therefore, instead of $v_k$ we can equivalently use $v_q\,q_k/q$ \cite{Catena:2021qsr} and obtain
\begin{align}
\label{eq:Flinear}
\mathcal{F}_{j_kn}(\vq,\vec v) &= \mathcal{F}_{\vec jn}^{\,q}(q,v,v_q)\,\frac{q_k}{q}\,,
\end{align}
where
\begin{align}
\mathcal{F}_{\vec jn}^{\,q}(q,v,v_q) &\equiv A_{\vec jn}(q,v,v_q)+v_q\,B_{\vec jn}(q,v,v_q)\,.
\end{align}
An analogous statement holds for $\mathcal{F}_{nj_k}$ as well. Similarly, for the contributions of type \labelcref{eq:Fjj}, the relevant%
\footnote{\label{ftnt:F55a}Again, the term proportional to $\epsilon_{k\ell m}\,q_m/m_e$ present in the explicit form of $\mathcal{F}_{j_{M_k}j_{M_{\ell}}}$ vanishes in the considered models. Moreover, the antisymmetric part of $\mathcal{F}_{j_{5_\ell}j_{5_m}}$, which is in principle non-zero, is irrelevant since it vanishes when contracted with $\Im\Delta\chi_{j_{5_k}^\dag j_{5_\ell}}$ which is proportional to $\delta_{k\ell}$, cf.~\cref{eq:chi55}.} coefficients $\mathcal{F}_{j_kj'_\ell}(\vq,\vec v)$ can be expressed as
\begin{align}
\label{eq:Fbilinear}
\mathcal{F}_{j_kj'_{\ell}}(\vq,\vec v) \equiv \mathcal{F}_{\vec j\vec j'}^{\,\delta}(q,v,v_q)\,\delta_{k\ell} + \mathcal{F}_{\vec j\vec j'}^{\,qq}(q,v,v_q)\,\frac{q_kq_{\ell}}{q^2}\,.
\end{align}
Consequently, the contributions given by \cref{eq:Fnn-nj-jj}, which always are scalar functions of $(q,v,v_q)$, can be expressed using
\begin{itemize}
\item the $\mathcal{F}_{n_0n_0}$ coefficient and the new scalar coefficients $\mathcal{F}_{\vec jn}^{\,q}$, $\mathcal{F}_{n\vec j}^{\,q}$, $\mathcal{F}_{\vec j\vec j'}^{\,\delta}$, $\mathcal{F}_{\vec j\vec j'}^{\,qq}$,
\item the $\chi_{n_0n_0}$ susceptibility and the contracted susceptibilities $\chi_{\vec j^\dag n}^q$, $\chi_{n^\dag \vec j}^q$, $\chi_{\vec j^\dag \vec j'}^\delta$, $\chi_{\vec j^\dag \vec j'}^{qq}$ given by
\begin{subequations}
\label{eq:contrsus}
\begin{align}
\chi_{\vec j^\dag n}^q(q,v_q)        &\equiv \sum_k\frac{q_k}{q}\,\chi_{j_k^\dag n}(\vec q, \vec v)\,,&
\chi_{n^\dag \vec j}^q(q,v_q)        &\equiv \sum_k\frac{q_k}{q}\,\chi_{n^\dag j_k}(\vec q, \vec v)\,,\\
\chi_{\vec j^\dag \vec j'}^\delta(q,v_q)  &\equiv \sum_{k\ell}\delta_{k\ell}\,\chi_{j_k^\dag j'_{\ell}}(\vec q, \vec v)\,,&
\chi_{\vec j^\dag \vec j'}^{qq}(q,v_q)    &\equiv \sum_{k\ell}\frac{q_kq_{\ell}}{q^2}\,\chi_{j_k^\dag j'_{\ell}}(\vec q, \vec v)\,.
\end{align}
\end{subequations}
\end{itemize}
Since the above contracted susceptibilities are just linear combinations of the susceptibilities $\chi_{a^\dag b}$, they also possess the analytic properties required to satisfy the Kramers-Kronig relations \labelcref{eq:KK-general,eq:KKgen2}.
For convenience, we also introduce the primed contracted susceptibilities, defined with the indices reversed:
\begin{subequations}
\label{eq:contrsusPrime}
\begin{align}
\chi'_{n_0n_0}                            &\equiv \chi_{n_0n_0}\,,&
\chi_{\vec j^\dag n}^{q\,\prime}          &\equiv \chi_{n^\dag \vec j}^q\,,&
\chi_{n^\dag \vec j}^{q\,\prime}          &\equiv \chi_{\vec j^\dag n}^q\,,&
\chi_{\vec j^\dag \vec j'}^{x\,\prime}    &\equiv \chi_{\vec j^{\prime\dag} \vec j}^x\,,
\end{align}
\end{subequations}
where $x=\delta,qq$.

In the following derivation of the theoretical upper bound, each $\mathcal{F}_i$ denotes either the $\mathcal{F}_{n_0n_0}$ coefficient or one%
\footnote{\label{ftnt:F55b}Strictly speaking, for the $\vec j_5\vec j_5$ contribution we consider a combination $\sum_{k\ell}\delta_{k\ell}\mathcal{F}_{j_{5_k}j_{5_\ell}} = 3\mathcal{F}_{\vec j_5\vec  j_5}^{\,\delta} + \mathcal{F}_{\vec j_5\vec j_5}^{\,qq}$. This is because $\chi_{j^\dag_{5_k}j_{5_\ell}}$ is proportional to $\delta_{k\ell}$, as can bee observed from \cref{eq:chi55}.}
of the scalar coefficients $\mathcal{F}(q,v,v_q)$ introduced on the right-hand sides of \cref{eq:Flinear,eq:Fbilinear}. Then, $\chi_i$ denotes, respectively, either $\chi_{n_0n_0}$ or the contracted susceptibility given by \cref{eq:contrsus} corresponding to $\mathcal{F}_i$. Using this notation, \cref{eq:rategenrecalled} can be rewritten as
\begin{align}
\Gamma &=  \frac{-i}{16 m_e^2 m_\chi^2} \,\sum_i \int \frac{{\rm d}\vec q}{(2 \pi)^3} \int {\rm d}\vec v \, f(\vec v)  \,
\mathcal{F}_i(q,v,v_q) \left(\chi_i - \chi^{\prime*}_i\right) (q,v_q)  \,.
\end{align}

We begin by performing the change of integration variables introduced in \cref{app:intdv} to obtain
\begin{align}
\label{eq:rategen2}
\begin{aligned}
\Gamma
&= \frac{-i}{32\,\pi^2 m_e^2 m_\chi^2}\sum_i
	\int_0^\infty {\mathrm d}q\,q^2\int_0^\infty {\mathrm d}\omega\,\left[
	\rho^{(0)}(\omega,q)\,\mathcal{F}_i^{(0)}(q,\omega,v)\right.\\
	&\mkern300mu+\left.
	\rho^{(2)}(\omega,q)\,\mathcal{F}_i^{(2)}(q,\omega,v)
	\right]
    \left(\chi_i - \chi^{\prime*}_i\right)(q,\omega)\,,
\end{aligned}
\end{align}
where
\begin{subequations}
\begin{align}
\rho^{(0)}(\omega,q)
= \frac{\pi}{q}\int_{v_q}^\infty {\mathrm d}v \, v \int_{-1}^1 {\mathrm d}\cos\alpha\,f(v,\cos\alpha)\,,\\
\rho^{(2)}(\omega,q)
= \frac{\pi}{q}\int_{v_q}^\infty {\mathrm d}v \, v^3 \int_{-1}^1 {\mathrm d}\cos\alpha\,f(v,\cos\alpha)\,.
\end{align}
\end{subequations}
Note that, though formally the integration in \cref{eq:rategen2} is performed over $0<q<\infty$ and $0<\omega<\infty$, its range is effectively limited by the DM velocity distribution $f$ to
\begin{subequations}
    \begin{align}
    0 &< q < q_{\max}\,,&
    q_{\max}&\equiv 2\,m_\chi\,(\vo + \vesc)\,,\\
    0 &< \omega < \omega_{\max}(q)\,,&
    \omega_{\max}(q)&\equiv q\,(\vo + \vesc) - \frac{q^2}{2\,m_\chi}\,,
    \end{align}
    \end{subequations}
corresponding to $v_q < v < \vo+\vesc$.
Moreover, the energy transfer $\omega$ is required to exceed the 1-electron ionization threshold, which subsequently imposes a threshold on the DM mass, as explained in results discussion in \cref{sec:numerical}.

Inspection of the explicit form of the functions $\mathcal{F}_{ab}$ given in \cref{app:F} shows that in general one has
\begin{align}
\label{eq:Fv}
\mathcal{F}_i(q,\omega,v) = \mathcal{F}_i^{(0)}(q,\omega) + v^2 \mathcal{F}_i^{(2)}(q,\omega)  \,,
\end{align}
where we made explicit that $\mathcal{F}_i(q,\omega,v)$ is either independent of $v$, or it depends on $v$ quadratically, cf. \cref{eq:Bv}. For readability, let us now keep only the $\mathcal{F}_i^{(0)}$ term. Calculation for the other term, which will be restored at the very end, follows analogously.

The final result for $\Gamma$ must be real, and since all $\mathcal{F}^{(0)}_i$'s are real in the considered models (as can be explicitly checked basing on \cref{app:F}), the part of the integral containing the real part of $\left(\chi_i - \chi^{\prime*}_i\right)$ must vanish. We thus obtain
\begin{align}
\label{eq:rategenFi}
\Gamma
&= \frac{1}{32 \pi^2 m_e^2 m_\chi^2}\,\sum_i
	\int_0^{q_{\max}} {\mathrm d}q\,q^2\int_0^{\omega_{\max}(q)} {\mathrm d}\omega\,\rho^{(0)}(\omega,q)\,
	\mathcal{F}^{(0)}_i(q,\omega,v)\,
	\Im\Delta\chi_i(q,\omega)\,,
\end{align}
where
\begin{align}
\Delta\chi_i\equiv\chi_i + \chi'_i\,,
\end{align}
such that $\Im\Delta\chi_i=\Im\,(\chi_i - \chi^{\prime*}_i)$. Note that, in the definition of $\Delta\chi_i$, none of the susceptibilities is complex-conjugated, so that, similarly to each $\chi_{a^\dag b}$, every $\Delta\chi_i(q,\omega)$ (treated as a function of $\omega$) is analytic on the complex upper half-plane. Therefore, it also obeys the Kramers-Kronig relations.

To obtain the theoretical upper bound, we perform two consecutive estimations.
Firstly, after dividing and multiply the integrand by $1=\omega/\omega$ for further convenience, we replace the value of $\omega\,\rho^{(0)}(\omega,q)\,\mathcal{F}^{(0)}_i(q,\omega)$ by its maximum for a given $q$:
\begin{align}\label{ineq:a}
\begin{aligned}
\Gamma
&\le \frac{1}{32 \pi^2 m_e^2 m_\chi^2}\,\sum_i
	\int_0^{q_{\max}} {\mathrm d}q\,q^2\,\max_\omega\left[
	\omega\,\rho^{(0)}(\omega,q)\,\mathcal{F}^{(0)}_i(q,\omega)
	\right]\times\\
	&\mkern300mu\times\int_0^{\omega_{\max}(q)} \frac{{\mathrm d}\omega}{\omega}\,
	\Im\Delta\chi_i(q,\omega)\,.
\end{aligned}
\end{align}
If, for a given value of $i$, $\Im\Delta\chi_i(q,\omega)$ is non-negative for each $\omega$, the above estimation \labelcref{ineq:a} is (for this particular $i$) a~trivial consequence of the fact that
    \begin{align}
    \omega\,\rho^{(0)}(\omega,q)\,\mathcal{F}^{(0)}_i(q,\omega)\,\Im\Delta\chi_i(q,\omega)
    \le
    \max_\omega\left[\omega\,\rho^{(0)}(\omega,q)\,\mathcal{F}^{(0)}_i(q,\omega)\right]\Im\Delta\chi_i(q,\omega)\,.
    \end{align}
This is not always the case; for instance, for $q = 15\text{ keV}$ the value of the $\Im\Delta\chi^\delta_{\vec j^\dag_M\vec j_M}(q,\omega)$ coupling (relevant for the anapole model) calculated for Si is negative for each $\omega$. However, as we checked numerically, for each considered coupling and material, the integral over momenta is dominated by those values of $q$ for which $\Im\Delta\chi_i$ is predominantly positive, so that \labelcref{ineq:a} is satisfied.

The second step makes use of the following inequality:
    \begin{align}\label{ineq:b}
    \int_0^{\omega_{\max}(q)}\frac{{\mathrm d}\omega}{\omega}\,\Im\Delta\chi_i(q,\omega)
    \le \pi\,U(q)^{-1}\,.
    \end{align}
To justify the above estimation for positive $\Im\Delta\chi_i$, we first extend the integration range:
    \begin{align}\label{ineq:b1}
    \int_0^{\omega_{\max}(q)}\frac{{\mathrm d}\omega}{\omega}\,\Im\Delta\chi_i(q,\omega)
    \le \int_0^\infty\frac{{\mathrm d}\omega}{\omega}\,\Im\Delta\chi_i(q,\omega)\,,
    \end{align}
and then, using the Kramers-Kronig relation \labelcref{eq:KKgen2}, replace the integral on the right-hand side by the value at $\omega=0$:
    \begin{align}
    \int_0^{\omega_{\max}(q)}\frac{{\mathrm d}\omega}{\omega}\,\Im\Delta\chi_i(q,\omega)
    \le \frac{\pi}{2}\,\Delta\chi_i(q,0)\,.
    \end{align}
Eventually, inequality \labelcref{ineq:b} is obtained by using the estimation
    \begin{align}\label{ineq:b2}
    \Delta\chi_i(q,0) \le 2\,U(q)^{-1}\,,
    \end{align}
corresponding for $\chi_i=\chi_{n_0n_0}$ to $\epsilon_r^{-1} > 0$, which should be the case for most materials relevant for DM detection (see an extended discussion of this issue in \cite{Lasenby:2021wsc}).

Neither condition \labelcref{ineq:b1} nor \labelcref{ineq:b2} is always satisfied; for example, $\labelcref{ineq:b1}$ does not hold in the aforementioned case of $\Im\Delta\chi^\delta_{\vec j^\dag_M\vec j_M}(q=15\text{ keV},\omega)$ for Si, and $\labelcref{ineq:b2}$ is broken by a~factor of ca.~20 for the $\Delta\chi^\delta_{\vec j^\dag_5\vec j_5}$ coupling (relevant in the anapole and the magnetic dipole model) calculated for the same material at $q=500\text{ eV}$. However, numerical tests confirm that, for all considered materials and couplings, inequality \labelcref{ineq:b} is always fulfilled. The reason is that, for any coupling and material, at most one of inequalities \labelcref{ineq:b1,ineq:b2} is broken at a time. If the first of them is fulfilled, the estimation \labelcref{ineq:b1} is very conservative for positive $\Im\Delta\chi_i$'s, so that the final conclusion holds even though \labelcref{ineq:b2} is incorrect. Conversely, if \cref{ineq:b1} does not hold, it means that $\Im\Delta\chi_i$ is predominantly negative, so the integral in \cref{ineq:b} can be limited from above by any positive value and one does not have to use \labelcref{ineq:b1} and the Kramers-Kronig relations to draw the final conclusion.

Using estimation \labelcref{ineq:b} we obtain the following theoretical upper bound on the interaction rate:
\begin{align}
\Gamma
&\le \frac{1}{32 \pi m_e^2 m_\chi^2}\,\sum_i
	\int_0^{q_{\max}} {\mathrm d}q\,q^2\,\max_\omega\left[
	\omega\,\rho^{(0)}(\omega,q)\,\mathcal{F}^{(0)}_i(q,\omega)
	\right]
	U(q)^{-1}\,.
\end{align}

In the general case of $\mathcal{F}_i(q,\omega,v) = \mathcal{F}^{(0)}_i(q,\omega) + v^2 \mathcal{F}^{(2)}_i(q,\omega)$ (cf. \cref{eq:Fv}), we apply the same procedure and find:
\begin{align}
\label{eq:ul2}
\begin{aligned}
\Gamma&\le
\Gamma_\text{opt}\equiv
	\frac{1}{32 \pi m_e^2 m_\chi^2}\,\sum_i
	\int_0^{q_{\max}} {\mathrm d}q\,q^2\,\max_\omega\left[
	\omega\,\rho^{(0)}(\omega,q)\,F^{(0)}_i(q,\omega)\right.\\
    &\mkern350mu+\left.
    \omega\,\rho^{(2)}(\omega,q)\,F^{(2)}_i(q,\omega)
	\right]
	U^{-1}(q)\,.
\end{aligned}
\end{align}
\Cref{eq:ul2} gives the theoretical upper bound on $\Gamma$ we find from the Kramers-Kronig relations. It is worth noting that due to the absence of $\chi$ in the final result, this bound is independent of the choice of the detector material. Thus, a comparison of the actual interaction rate \labelcref{eq:rategenFi} with the result of \cref{eq:ul2} provides a way to evaluate the given material as a potential target in direct detection experiments. Below, we apply this general bound to the specific case of anapole, magnetic dipole, and electric dipole DM.

\subsubsection{DM of spins different than 1/2}
In our analysis, we assume the dark particles to have spin 1/2. Effective theory of interactions of DM of spin up to $1$ with nucleons or electrons has been the topic of numerous works, including \cite{Gondolo:2021fqo,Gondolo:2020wge,Liang:2024ecw}. As can be observed from Table~1 of \cite{Liang:2024ecw}, to describe interactions between spin-1 DM and electrons one must employ more types of effective operators than in the fermion DM case. This would affect our calculations in two ways:
    \begin{enumerate}
    \item coefficients $F_{ab}$ for $a,b=n_0,n_A,\vec j_5,\vec j_M,\vec j_E$, given by \cref{eq:F,eq:F2App,eq:F2AppCont}, would contain more terms coming from the new operators depending on the generalized densities and currents analysed by us;
    \item some of the effective operators relevant for spin-1 DM cannot be expressed in the form of our \cref{eq:Vefftot}, so that including them would require indices $a$ and $b$ to cover a broader range of generalized densities and currents. For instance, operator $\mathcal{O}_{21}$ defined in \cite{Liang:2024ecw} requires introducing in \cref{eq:Vefftot} a term proportional to
        \begin{align}
        \frac{i}{2m_e}\,
        \left[
        \overleftarrow\nabla_{\vec r_e}
        \cdot\vec{\mathcal{S}}_X^\text{sym}
        \cdot\vec\sigma_e\,e^{i\vec q\cdot\vec r_e}
        -
        e^{i\vec q\cdot\vec r_e}\,\vec\sigma_e
        \cdot\vec{\mathcal{S}}_X^\text{sym}
        \cdot\overrightarrow\nabla_{\vec r_e}
        \right]\;,
        \end{align}
    with $(\mathcal{S}_X^\text{sym})^{ij}\equiv \frac12(S_X^iS_X^j+S_X^jS_X^i)$, where $\vec S_X$ denotes a vector formed by the spin operators of the vector DM particle. As can be observed, such a term cannot be decomposed into a DM-related and an $e^-$-related part in any other way than by introducing a generalized ``electronic coupling tensor'' $\vec{\mathcal J}$ given by
        \begin{align}
        \mathcal J^{ij}\equiv\frac{i}{2m_e}\,
        \left[
        \overleftarrow\partial^i_{\vec r_e}
        \sigma_e^j\,e^{i\vec q\cdot\vec r_e}
        -
        e^{i\vec q\cdot\vec r_e}\,
        \overrightarrow\partial^j_{\vec r_e}
        \sigma_e^i
        \right]\;.
        \end{align}
    \end{enumerate}
Consequently, analysis of the vector DM case requires adding new contributions to the $\mathcal{F}_i$ terms both in \cref{eq:rategenFi}, which expresses the actual DM-$e^-$ interaction rate, and in \cref{eq:ul2}, which provides the theoretical upper bound. The ratio $\Gamma_\text{opt}/\Gamma$ would depend on the new terms in a non-trivial way, so that it is difficult to predict to which extent our results would apply to the general vector DM case.
Note that all the new contributions to the matrix element squared would be expressible in terms of the generalized susceptibilities employed in this work; no new generalized susceptibilities would be needed \cite{Liang:2024ecw}.

On the contrary, effective operators required to describe interactions of scalar DM with electrons form a subset of those employed in the fermion DM case. Therefore, adjusting our calculations to the spin-0 case would be straightforward, as it only requires dropping terms specific to the spin-1/2 case. However, precise values of the final ratio between the theoretical optimum and the actual interaction rate cannot be predicted prior to an actual calculation.

Regardless of the aforementioned differences in detailed forms of expressions and numerical results, the methodology we have used remains valid for any spin of DM particles, so that it can be straightforwardly extended and used to generalize our approach to the cases of DM of spins other than 1/2.

\subsection{Interaction rate in specific effective models}
\label{sec:analytic_models}
\subsubsection{Anapole}
\label{sec:anapole}
DM has an anapole moment if it couples to the photon via the higher order electromagnetic coupling,
\begin{align}
\mathscr{L}=\frac{1}{2}\frac{g}{\Lambda^2} \bar{\chi} \gamma^\mu \gamma^5 \chi\, \partial^\nu F_{\mu\nu} \,,
\label{eq:anapole}
\end{align}
where $F_{\mu\nu}$ is the electromagnetic field strength tensor, $\chi$ is a Majorana four-component spinor describing the DM particle, $g$ is a coupling constant while $\Lambda$ is an energy scale. In the non-relativistic limit, \cref{eq:anapole} implies the following amplitude for DM scattering by a free electron~\cite{Catena:2019gfa},
\begin{align}
\mathscr{M}=\frac{4 e g}{\Lambda^2} m_\chi m_e \left[
    2 \left(\vec v_{\rm el}^\perp \cdot \xi^{\dagger s'} \vec S_\chi \xi^s \right) \delta^{r' r} +
    g_e \left( \xi^{\dagger s'} \vec S_\chi \xi^s \right) \cdot \left( i\frac{\vec q}{m_e} \times \xi^{\dagger r'} \vec S_e \xi^r  \right) 
    \right]\,,
\label{eq:Manapole}
\end{align}
where $\vec S_{\chi}\equiv\vec\sigma_{\chi}/2$ $(\vec S_e\equiv\vec\sigma_e/2)$ denotes the spin matrix corresponding to the dark particle (electron), $g_e=2$, and $\vec v_\text{el}^\perp$ is the so-called transverse relative velocity, which is the component of the relative DM-$e^-$ velocity transverse to the momentum transfer in the case of elastic scattering.
The associated transition potential depends on the electron density, paramagnetic current, and spin current~\cite{Catena:2024rym}:
\begin{align}
V^{ss'}_{\rm eff} = -&\frac{1}{4m_e m_\chi V} \left[F_0^{ss'} n_0(-\vec q)
+ \vec F_5^{ss'}\cdot \vec j_5(-\vec q) 
+ \vec F_M^{ss'}\cdot \vec j_M(-\vec q) 
\right] \,.
\label{eq:Veffa}
\end{align}
where the functions $F_0^{ss'}$, $\vec F_5^{ss'}$ and $\vec F_M^{ss'}$ are listed in \cref{eq:F} with $c_8$ and $c_9$ explicitly given by
\begin{subequations}
    \label{eq:ca}
\begin{align}
    c_8 &= 8 e m_e m_\chi\frac{g}{\Lambda^2}\, ,\\
    c_9 &= -8 e m_e m_\chi\frac{g}{\Lambda^2}\, ,
\end{align}
\end{subequations}
and all other coupling constants set to zero. By combining \cref{eq:ca,eq:F2App,eq:rategenFi}, we find
\begin{align}
\label{eq:rategen_a}
\Gamma =  \frac{1}{32\pi^2 m_e^2 m_\chi^2} \,\sum_i \int_0^{q_{\max}}{\mathrm d}q\,q^2\int_0^{\omega_{\max}(q)}{\mathrm d}\omega\,
    \Im\Delta\chi_i(q,\omega)\,
    &\Big[\rho^{(0)}(\omega,q)\, \mathcal{F}^{(0)}_i(q,\omega)\\
    &\quad+\rho^{(2)}(\omega,q)\, \mathcal{F}^{(2)}_i(q,\omega)\Big]\notag\,,
\end{align}
with $\mathcal{F}_i^{(0)}$, $\mathcal{F}_i^{(2)}$ defined in \cref{eq:Fv}, and the non-zero $\mathcal{F}_i$ coefficients being
\begin{subequations}
\label{eq:Fa}
\begin{align}
\mathcal{F}_{n_0 n_0}(q, \omega, v)                         &= \frac14\,\left(v^2-\frac{\omega}{2\,m_\chi}-\frac{q^2}{4\,m_\chi^2}\right) \, c_8^2 \,,\\
3\,\mathcal{F}_{\vec j_5\vec j_5}^{\,\delta}(q, \omega, v) + \mathcal{F}_{\vec j_5\vec j_5}^{\,qq}(q, \omega, v)
                                                            &=\sum_k\mathcal{F}_{j_{5_k} \, j_{5_k}}(q, \omega, v)\\
                                                            &=\frac18\,\frac{q^2}{m_e^2}\, c_9^2\,,\notag\\
\mathcal{F}_{\vec j_M\vec j_M}^{\,\delta}(q, \omega, v)         &=\frac14\,c_8^2\,,\\
\mathcal{F}_{\vec j_M n_0}^{\,q}(q, \omega, v)
= \mathcal{F}_{n_0 \vec j_M}^{\,q}(q, \omega, v)                &= - \frac14\,\frac{\omega}{q}\, c_8^2\,.
\end{align}
\end{subequations}

\subsubsection{Magnetic dipole}
\label{sec:magdip}
The Lagrangian for the magnetic dipole coupling between a Dirac DM field $\psi$ and the photon field $A_\mu$ is
\begin{align}
\mathscr{L}&= \frac{g}{\Lambda} \, \bar{\psi}\sigma^{\mu\nu}\psi \, F_{\mu\nu}\,, 
\end{align}
where $F_{\mu\nu} = \partial_\mu A_\nu -  \partial_\nu A_\mu$, and $\sigma^{\mu \nu}= i[\gamma^\mu,\gamma^\nu]/2$. The associated amplitude for DM-electron scattering is
\begin{align}
\begin{aligned}
\mathscr{M}&=
\frac{e g}{\Lambda}  \Bigg\{
    4m_e\delta^{s's}\delta^{r'r} +\frac{16m_\chi m_e}{q^2}  i\vec q \cdot \left(\vec v_\text{el}^\perp \times \xi^{\dagger s'} \vec S_\chi \xi^s \right)\delta^{r' r} \\
    &-  \frac{8 g_em_\chi}{q^2} \Bigg[\left( \vec q \cdot \xi^{\dagger s'} \vec S_\chi \xi^s \right)\left( \vec q \cdot \xi^{\dagger r'} \vec S_e \xi^r \right)
    - q^2  \left( \xi^{\dagger s'} \vec S_\chi \xi^s \right)\cdot \left( \xi^{\dagger r'} \vec S_e \xi^r \right)
    \Bigg] \Bigg\}\,,
\end{aligned}
\end{align}
with $\vec S_{\chi,e}$ and $\vec v_\text{el}^\perp$ defined as in \cref{eq:Manapole}.
This amplitude implies an effective transition potential of the same form as in \cref{eq:Veffa}, but the only coupling constants that are now different from zero are
\begin{subequations}
    \label{eq:cm}
\begin{align}
    c_1 &= 4 e m_e \frac{g}{\Lambda}\, ,\\
    c_4 &= 16 e m_\chi\frac{g}{\Lambda}\, ,\\
    c_5 &= \frac{16em_e^2 m_\chi}{q^2} \frac{g}{\Lambda}\, ,\\
    c_6 &= -\frac{16em_e^2 m_\chi}{q^2} \frac{g}{\Lambda}\, .
\end{align}
\end{subequations}
By combining \cref{eq:cm,eq:F2App,eq:rategenFi}, we obtain a transition rate formula with the same structure as in \cref{eq:rategen_a}, but the non-vanishing $\mathcal{F}_i$ coefficients are now given by
\begin{subequations}
\label{eq:Fm}
\begin{align}
\mathcal{F}_{n_0 n_0}(q, \omega, v)                 &= c_1^2 + \frac{q^2\,v^2-\left(\omega+\frac{q^2}{2\,m_\chi}\right)^2}{4\,m_e^2}\, c_5^2\,,\\
3\,\mathcal{F}_{\vec j_5\vec j_5}^{\,\delta}(q, \omega, v) + \mathcal{F}_{\vec j_5\vec j_5}^{\,qq}(q, \omega, v)
                                                    &= \sum_k\mathcal{F}_{j_{5_k} \, j_{5_k}}(q, \omega, v)\\
                                                    &= \frac{3}{16} c_4^2 + \frac{q^4}{16 m_e^4} c_6^2 + \frac{q^2}{8 m_e^2}c_4c_6 \,,\notag\\
\mathcal{F}_{\vec j_M\vec j_M}^{\,\delta}(q, \omega, v) &=\frac{q^2}{4 m_e^2}\, c_5^2\,,\\
\mathcal{F}_{\vec j_M \vec j_M}^{\,qq}(q, \omega, v) &=-\frac{q^2}{4 m_e^2}\, c_5^2\,.
\end{align}
\end{subequations}

\subsubsection{Electric dipole}
\label{sec:eldip}
The case of electric dipole DM is characterized by the Lagrangian 
\begin{align}
\mathscr{L}_{\rm electric}&= \frac{g}{\Lambda} \, i\bar{\psi} \sigma^{\mu\nu} \gamma^5 \psi \, F_{\mu\nu} \,.
\end{align}
The associated amplitude for DM-electron scattering takes the form 
\begin{align}
\mathscr{M}= \frac{e g}{\Lambda} \frac{16 m_\chi m_e}{q^2} i\vec{q} \cdot \left( \xi^{\dagger s'} \vec{S}_\chi \xi^s \right)\delta^{r'r}\,.
\end{align}
This implies that the total transition rate $\Gamma$ can be written as
\begin{align}
\label{eq:rategen_e}
\Gamma =  \frac{1}{16\pi^2 m_e^2 m_\chi^2} \,\int_0^{q_{\max}}{\mathrm d}q\,q^2\int_0^{\omega_{\max}(q)}{\mathrm d}\omega\,
    \rho^{(0)}(\omega,q)\, \mathcal{F}_{n_0n_0}(q,\omega)\,\Im\chi_{n_0n_0}(q,\omega)\,,
\end{align}
with 
\begin{align}
\mathcal{F}_{n_0 n_0}(q, \omega) = \frac{q^2}{4 m_e^2} c_{11}^2\,,
\label{eq:Fe}
\end{align}
where
\begin{align}
    c_{11} &= \frac{16 e m_\chi m_e^2}{q^2}\frac{g}{\Lambda}\,.
\end{align}

\subsection{Numerical results}
\label{sec:numerical}
In order to numerically evaluate the transition rate formulae in \cref{eq:rategen_a,eq:rategen_e}, and the associated theoretical upper bound in \cref{eq:ul2}, we use the relations between the relevant generalized susceptibilities of~\cite{Catena:2024rym}, $\chi_{a^\dagger b}$, and the atomic and crystal responses (i.e., $W$ functions) derived in~\cite{Catena:2019gfa}, to express the contracted susceptibilities defined in \cref{eq:contrsus} as
\begin{subequations}
\label{eq:chi}
\begin{align}
\Im \chi_{n_0n_0}(q,\omega)
&= \frac{\Im\Sigma_{n_0n_0}(q,\omega)}
{\left|1 + U(q)\,[1-G(q)]\,\Sigma_{n_0n_0}(q,\omega)\right|^2}\,,\\
\Im \Delta \chi_{\vec j_5^\dag \vec j_5}^{qq}(q,\omega)
&=\frac{1}{3}\,\Im \Delta \chi_{\vec j_5^\dag \vec j_5}^{\delta}(q,\omega)
= 2\,\Im\Sigma_{n_0n_0}(q,\omega)\,,\\
\Im \Delta \chi^\delta_{\vec j_M^\dag \vec j_M}(q,\omega)
&= \sum_{k\ell}\delta_{k\ell}\,\Im \Delta \chi_{j_{M_k}^\dagger j_{M_{\ell}}}(q,\omega)\notag\\
&= \frac{2\pi^2 \widetilde{\Omega}}{\omega} \bigg[ \frac{q^2}{4m_e^2} W_{1}(q,\omega) + W_{3}(q,\omega) + \Re W_{2}(q,\omega) \bigg]\\
&\quad- 2\,\frac{\omega^2}{q^2}\left[\Im\Sigma_{n_0n_0}(q,\omega) - \Im\chi_{n_0n_0}(q,\omega)\right] \,,\notag\\
\Im \Delta \chi^{qq}_{\vec j_M^\dag \vec j_M}(q,\omega)
&= \sum_{k\ell}\frac{q_k q_{\ell}}{q^2}\,\Im \Delta \chi_{j_{M_k}^\dagger j_{M_{\ell}}}(q,\omega)\notag\\
&=\frac{2\pi^2 \widetilde{\Omega}}{\omega} \bigg[ \frac{q^2}{4\,m_e^2} W_{1}(q,\omega) + \frac{m_e^2}{q^2}\,W_{4}(q,\omega) + \Re W_{2}(q,\omega) \bigg]\\
&\quad- 2\,\frac{\omega^2}{q^2}\left[\Im\Sigma_{n_0n_0}(q,\omega) - \Im\chi_{n_0n_0}(q,\omega)\right]\,,\notag\\
\Im \Delta \chi^q_{\vec j_M^\dag n_0}(q,\omega)
&= \sum_k\frac{q_k}{q}\,\Im \Delta \chi_{j_{M_k}^\dag n_0}(q,\omega)\notag\\
&=\frac{m_e}{q}\,\frac{\pi^2 \widetilde{\Omega}}{\omega} \bigg[ \frac{q^2}{m_e^2} W_{1}(q,\omega) + 2\,\Re W_{2}(q,\omega) \bigg]
\label{eq:chiM0}\\
&\quad-2\,\frac{\omega}{q}\left[\Im\Sigma_{n_0n_0}(q,\omega) - \Im\chi_{n_0n_0}(q,\omega)\right]\,.\notag
\end{align}
\end{subequations}
The result for $\Im\Delta\chi^q_{n_0\vec j_M}$ is the same as the one obtained for $\Im\Delta\chi^q_{\vec j_M^\dag n_0}$, given by \cref{eq:chiM0}. Let us also notice that before contraction, the $\Im\Delta\chi_{j_{5_k}^\dag j_{5_l}}$ susceptibility is proportional to the delta function \cite{Catena:2024rym}:
\begin{align}
\label{eq:chi55}
\Im \Delta \chi_{j_{5_k}^\dagger j_{5_{\ell}}}(q,\omega)
&= 2\,\delta_{k\ell} \,\Im\Sigma_{n_0n_0}(q,\omega)\,,
\end{align}
which was important for the reasoning of \cref{ftnt:F55a,ftnt:F55b}.
In \cref{eq:chi,eq:chi55}, we express the in-medium corrections via the density-density response function $\Sigma_{n_0 n_0}$ and the local-field factor $G$~\cite{Catena:2024rym}.
Our derivation assumes that the material is non-spin-polarized, which allows for simplifications in the electron spin sums, resulting in the presented relation between the susceptibilities and the electron number density. For non-spin-polarized materials, $\Sigma_{j_\alpha n_0}$ and $\Sigma_{n_0j_\beta}$ are different from zero only for $j_\alpha,j_\beta\in\{j_M,n_0\}$ \cite{Catena:2024rym}.
The relation between $\Sigma_{n_0n_0}$ and the numerical data is explained below. $\widetilde{\Omega}$ is defined so that  $\widetilde{\Omega} V/M=1/\widetilde{m}$, where $M$ is the detector mass and $\widetilde{m}$ is the unit cell mass in the case of crystals, and the argon or xenon atom mass in the case of noble liquids~\cite{Catena:2024rym}.\footnote{In the case of crystals, $\widetilde{\Omega} = V_\text{cell}^{-1}$.} This implies that the total electronic transition rate per unit detector mass induced by  $n_\chi V$ DM particles, i.e., $n_\chi V \Gamma/M$, is independent of $V$. Here, $n_\chi$ is the local DM number density.

For the functions $\Re  W_{2}$, $W_{3}$, and $W_{4}$, we use values that were tabulated in~\cite{Catena:2019gfa} with {\sffamily DarkART}~\cite{DarkART}%
\footnote{The $\Re  W_{2}$ function computed by {\sffamily DarkART} has to be multiplied by a factor of minus $1$ to account for a missing minus sign in the definition of the vectorial form factor $\vec f_{i\rightarrow f}$ in~\cite{Catena:2019gfa}. The correct definition for $\vec f_{i\rightarrow f}$  is
\begin{align}
\vec f_{i\rightarrow f}(\vec q) = -\frac{i}{m_e} \int {\rm d}\vec r\, \psi^*_{f}(\vec r)e^{i \vec q \cdot \vec r} \nabla_{\vec r}\psi_i(\vec r) \,,
\end{align}
where $\psi_i$ and $\psi_f$ are the initial and final electronic wave functions. Notice also that {\sffamily DarkART} computes all $W$'s as a function of $q$ and $k'$, where $k'$ is the asymptotic momentum of the final state electron. Following~\cite{Catena:2022fnk}, we use
\begin{align}
W_j(q,\omega) \equiv \frac{\omega}{2 \pi} \sum_{n\ell} \int \frac{{\rm d}k'}{k'} \, W_j^{n\ell}(k',q) \, \delta(\omega - E_{k'\ell' m'} + E_{n\ell m})\,,\qquad j=1,2,3,4 \,,
\label{eq:newold}
\end{align}
to convert the $W_j^{n \ell}(q,k')_i$ functions of~\cite{Catena:2019gfa} to functions of $q$ and $\omega$, i.e., $W_j(q,\omega)$. Here, $E_f=E_{k'\ell' m'}$ and $E_i=E_{n\ell m}$.
}
for argon and xenon detectors, and in~\cite{Catena:2021qsr} with {\sffamily QEdark-EFT}~\cite{Urdshals2021May} for germanium and silicon targets. The $W_1$ function is related to the imaginary part of the density-density response function, $\Sigma_{n_0n_0}$, in the following way:
\begin{align}\label{eq:W1}
\Im\Sigma_{n_0 n_0}(q,\omega)=\frac{\pi^2\,\widetilde\Omega}{\omega}\,W_1(q,\omega)\,.
\end{align}
For crystal detectors, we extract the density-density response function $\Sigma_{n_0 n_0}$ from the equation
\begin{align}
\Sigma_{n_0 n_0}(q,\omega)=U(q)^{-1}\left[\epsilon_r^{\rm GPAW}(q,\omega)-1\right]\,,
\end{align}
where the dielectric function $\epsilon_r^{\rm GPAW}$ was calculated in~\cite{Knapen:2021run} using GPAW~\cite{Mortensen_2024}, and implemented in {\sffamily DarkELF}~\cite{Knapen:2021bwg}. The $W_1$ function is then obtained from \cref{eq:W1}. Conversely, in the case of argon and xenon, we use the values of $W_1$ tabulated in~\cite{Catena:2021qsr} to calculate the imaginary part of $\Sigma_{n_0n_0}$. The real part of $\Sigma_{n_0n_0}$ can be obtained from the imaginary one by applying the general Kramers-Kronig relation given by \cref{eq:KK-general}.\footnote{$\Sigma_{n_0n_0}$ satisfies all assumptions of \cref{eq:KK-general} as well as $\chi_{n_0n_0}$.}
For $G$ we use the expressions in Eq.~(94) of~\cite{Catena:2024rym} for germanium and silicon, while we set $G=0$ for argon and xenon.

\begin{figure}[t]\begin{center}
\includegraphics[width=0.6\textwidth]{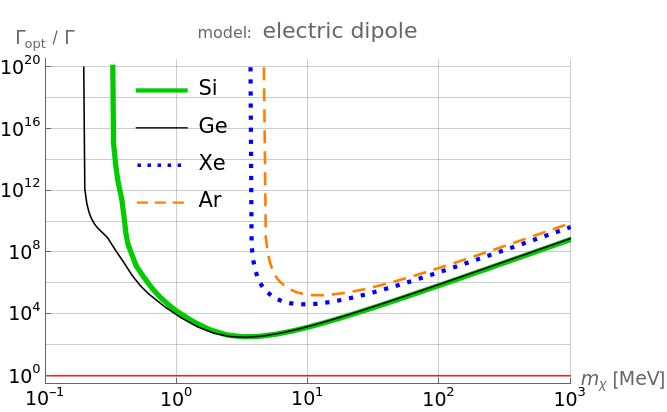}
\caption{Ratio between the theoretical upper bound $\Gamma_\text{opt}$ and the actual value of the interaction rate $\Gamma$, obtained within the electric dipole model, as a function of DM mass. The lines correspond to different materials: silicon (thick green), germanium (thin black), xenon (dotted blue), and argon (dashed orange). The horizontal red line marks the value of 1, for which the actual interaction rate would saturate the bound.}
\label{fig:el_dipole}
\end{center}\end{figure}

\begin{figure}[t]\begin{center}
\includegraphics[width=0.6\textwidth]{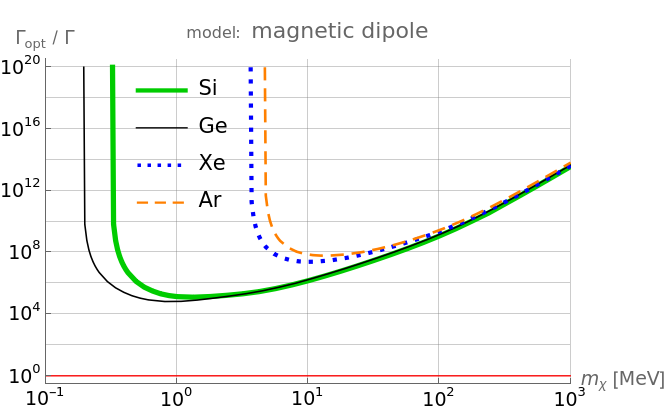}
\caption{As in \cref{fig:el_dipole}, but for the magnetic dipole model.}
\label{fig:mag_dipole}
\end{center}\end{figure}

\begin{figure}[t]\begin{center}
\includegraphics[width=0.6\textwidth]{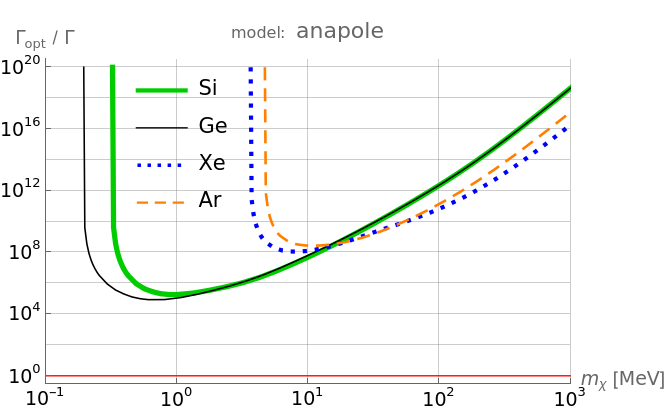}
\caption{As in \cref{fig:el_dipole,fig:mag_dipole}, but for the anapole model.}
\label{fig:anapole}
\end{center}\end{figure}

\Cref{fig:el_dipole,fig:mag_dipole,fig:anapole} show the ratio between the optimal total interaction rate $\Gamma_\text{opt}$, given by \cref{eq:ul2}, and the actual total interaction rate $\Gamma$, obtained for a given material from \cref{eq:rategenFi}. The results presented in \cref{fig:el_dipole} have been obtained in the electric dipole model, described briefly in \cref{sec:eldip}, while \cref{fig:mag_dipole,fig:anapole} correspond to the magnetic dipole (\cref{sec:magdip}) and anapole (\cref{sec:anapole}) models, respectively. In each plot, the thick green line represents the values obtained for silicon, the thin black line corresponds to germanium, the dotted blue line denotes the results for xenon and the dashed orange line has been obtained for argon. The horizontal red line marks the reference value of $1$, for which the actual interaction rate $\Gamma$ would saturate the theoretical upper bound $\Gamma_\text{opt}$. Note that $\Gamma_\text{opt}$ does not depend on the chosen material, so the plotted ratio can serve as an evaluation of a given material's quality with respect to its usefulness in direct searches for DM.

The maximal energy transfer from a dark particle of mass $m_\chi$, which corresponds to a total absorption of its energy, is $m_\chi\,(v_\oplus+v_\text{esc})^2\,/\,2$, where $v_\oplus+v_\text{esc}\approx 800~\text{m/s}$ is the maximal velocity of the dark particle according to the standard halo model (see \cref{app:vDM}). For the materials considered in our manuscript, excitations caused by DM particles of masses lower than $0.1~\text{MeV}$ would fail to exceed the 1-electron ionization threshold. In such a case, the detection method should be focused on different channels, such as phononic excitations. This can clearly be observed in our plots, where the ratio between the optimum and the actual rate tends to infinity as DM mass approaches the value corresponding to the 1-electron ionization threshold: $m_\chi = 0.19~\text{MeV}$ for Ge, $m_\chi = 0.32~\text{MeV}$ for Si, $m_\chi = 4.6~\text{MeV}$ for Ar, $m_\chi = 3.6~\text{MeV}$ for Xe.

It can be observed that, for Si and Ge detectors, the ratio approaches values around $10^2$ in the electric dipole model, while in the anapole and the magnetic dipole model, the ratio is always very large, i.e., above $10^4$. This is correlated with the fact that the response to the density-density coupling (the only one relevant for the electric dipole) is close to the maximal possible, while the response to the $\Delta\chi^\delta_{\vec j^\dag_M\vec j_M}$, $\Delta\chi^{qq}_{\vec j^\dag_M\vec j_M}$ couplings which dominate for the anapole and the magnetic dipole, is far from optimal. In contrast, for Ar and Xe detectors, our theoretical upper bound is at least four orders of magnitude larger than the actual rate, regardless of the model. For all materials and models considered in this work, the ratio reported in \Cref{fig:el_dipole,fig:mag_dipole,fig:anapole} is minimal for a value of the DM particle mass in the range between 0.5 and 15 MeV.

\section{Summary and conclusions}
\label{sec:conclusions}

We investigated the analytic properties of the rate of DM-induced electronic transitions in materials within a framework that combines an effective theory description of DM-electron interactions with linear response theory. Within this framework, the rate of DM scattering events in materials can be expressed in terms of generalized susceptibilities describing the response of detectors to an external DM perturbation. We found that the rate of DM-induced electronic transitions in materials admits a theoretical upper bound under general assumptions on the underlying DM-electron coupling. This bound applies to models where DM couples to the electron density as well as the electron spin, paramagnetic and Rashba currents. We obtained this result by applying a set of the Kramers-Kronig relations to the generalized susceptibilities used in~\cite{Catena:2024rym} to express the rate of DM-induced electronic transitions in materials.   

We evaluated our theoretical upper bound numerically for Ar, Xe, Ge and Si targets and found that, while Ge and Si detectors are generically closer to saturate this theoretical upper bound, they are still far from saturation, unless DM couples to the electron density. This motivates the exploration of different classes of materials to effectively
probe such coupling forms.

\acknowledgments 
RC and MI acknowledge support from an individual research grant from the Swedish Research Council (Dnr.~2022-04299). RC has also been funded by the Knut and Alice Wallenberg Foundation, and performed this research within the ``Light Dark Matter'' project (Dnr. KAW 2019.0080). MI has also been supported by the Royal Society of Arts and Sciences in Gothenburg travel grant (Dnr. 2025-963).

\appendix

\section{DM velocity distribution}
\label{app:vDM}
We assume that the distribution of DM velocities in the laboratory frame, $f(\vec v)$, has the form of truncated Maxwell-Boltzmann distribution,
\begin{align}
f(\vec v)&= \N\,\exp\left[-\frac{(\vv + \vvo)^2}{v_0^2}\right]\,\theta(\vesc-|\vv+\vvo|)\,,
\end{align}
where $\vvo$ is Earth's velocity relative to the galactic centre, $\vesc$ is the galactic escape velocity, and $v_0$ is the most probable DM speed (in the galactic reference frame, in which the mean DM velocity is zero). The normalization constant $\N$ is given by
\begin{align}
\N&= \frac{1}{2\pi\,v_0^3}\left(
\frac{\sqrt\pi}{2}\,\erf\left[\frac{\vesc}{v_0}\right] - \frac{\vesc}{v_0}\,\exp\left[-\frac{\vesc^2}{v_0^2}\right]
\right)^{-1}\,.
\end{align}
The values of $\vo\equiv\vvo$, $\vesc$ and $v_0$ adopted in this work after \cite{Baxter:2021pqo} are
\begin{align}
\vo   = 250.5\kms\,,\quad
\vesc = 544\kms\,,\quad
v_0   = 238\kms\,.
\end{align}

\section{Integration over velocities: from \texorpdfstring{${\rm d} \vec v$ to ${\rm d} \omega$}{dv to d omega}}
\label{app:intdv}

In general, the electronic transition rate in \cref{eq:rategen} can be written as
	\begin{align}
	\Gamma
	&= \int {\mathrm d}\vec q\,{\mathrm d}\vec v\,f(\vec v)\,B(\vec v, \vec q)\,,
	\label{eq:rateB}
	\end{align}
where $f(\vec v)$ describes the distribution of DM velocities in the laboratory frame (see \cref{app:vDM}), and $B(\vec v, \vec q)$ is a model dependent function we do not need to specify here. If we assume that the target material is isotropic (see \cref{app:isotropy}), then
	\begin{align}\label{eq:appendix:gamma}
	\Gamma
	&= \int {\mathrm d}\vec q\,{\mathrm d}\vec v\,f(\vec v)\,B(\omega, q, v)\,,
	\end{align}
where\footnote{To simplify the notation, we write $\omega$ instead of $\omega_{\vec v, \vec q}$, cf. \cref{eq:omega}.}
	\begin{align}
    \label{eq:omegaApp}
	\omega \equiv \vec q\cdot \vec v - \frac{q^2}{2\,m_\chi} \,.
	\end{align}
Let us work in a coordinate system where Earth's velocity with respect to the galactic centre (denoted by $\vvo$), DM velocity in the laboratory frame ($\vec v$), and momentum transfer ($\vec q$) are related by the following relations:
    \begin{subequations}
    \begin{align}
	\vvo	&= \vo\,\vec e_z\,,\\
	\vv		&= v\,R(\alpha,\beta)\,\vec e_z\,,\\
	\vec q	&= q\,R(\alpha,\beta)\,R(\theta,\phi)\,\vec e_z\,,
	\end{align}
    \end{subequations}
where
	\begin{align}
	R(\theta,\phi) &\equiv
		\mmm{\cos\phi}{-\sin\phi}{0}{\sin\phi}{\cos\phi}{0}{0}{0}{1}
		\mmm{\cos\theta}{0}{\sin\theta}{0}{1}{0}{-\sin\theta}{0}{\cos\theta}
	\end{align}
is a rotation matrix, so that
	\begin{align}
	\vv\cdot\vvo	&= v\vo\cos\alpha\,,&
	\vv\cdot\vq		&= vq\cos\theta\,.
	\end{align}
Next, we perform the change of integration variables $(\vec q,\,\vec v)\rightarrow (q,\,\omega,\,v,\,\alpha,\,\beta,\,\phi)$, such that
	\begin{align}
	{\mathrm d}\vec q\,{\mathrm d}\vec v
	&= q\,{\mathrm d}q\,{\mathrm d}\omega\,v\,{\mathrm d}v\,{\mathrm d}\cos\alpha\,{\mathrm d}\beta\,{\mathrm d}\phi\,,
	\end{align}
which, after integrating over the irrelevant angles $\beta$ and $\phi$ ($\Gamma$ does not depend on them), turns to
	\begin{align}\label{eq:appendix:variables}
	{\mathrm d}\vec q\,{\mathrm d}\vec v
	&= 4\pi^2\,q\,{\mathrm d}q\,{\mathrm d}\omega\,v\,{\mathrm d}v\,{\mathrm d}\cos\alpha.
	\end{align}
The integration range is
	\begin{align}
	0&<q<\infty\,,&
	0&<\omega<\infty\,,&
	v_q&<v<\infty\,,&
	-1&<\cos\alpha<1\,,
	\end{align}
where
	\begin{align}
	v_q \equiv \frac{\omega}{q} + \frac{q}{2\,m_\chi}\,,
	\end{align}
and the $\omega>0$ requirement has been introduced for physical reasons. Note that the velocity $v$ is bounded from above by the cut-off value $\vesc + \vo$, which is imposed via the step function in $f(\vv)$ (see \cref{app:vDM}) and affects the actual range of integration over $dq$ and $d\omega$. 

\subsection{Velocity-independent \texorpdfstring{$B$}{B}}
When the function $B$ introduced in \cref{eq:rateB} is independent of $v$, the rate $\Gamma$ can be written as follows:
	\begin{align}
	\Gamma
	= 4\pi\int_0^\infty q^2{\mathrm d}q\int_0^\infty {\mathrm d}\omega\,\rho^{(0)}(\omega,q)\,B(\omega,q)\,,
	\end{align}
where
	\begin{align}
	\rho^{(0)}(\omega,q)
	\equiv \frac{\pi}{q}\int_{v_q}^\infty v\,{\mathrm d}v\int_{-1}^1 {\mathrm d}\cos\alpha\,f(\vec v)\,.
	\end{align}
By analytically evaluating the angular and velocity integrals in the above equation, we find
	\begin{align}
    \begin{aligned}
	\rho^{(0)}(\omega,q)
		&= \frac{\N\,\pi\,v_0^2}{2\,q}\,\times\\
		&\mkern10mu\times\begin{cases}
		\frac{\sqrt\pi}{2}\,\frac{v_0}{\vo}\left[ \erf\left(x_{+}\right) + \erf\left(x_{-}\right) \right]
			-2\,e^{-z^2}
			& \text{for }
			v_q < \vesc - \vo\\
		\frac{\sqrt\pi}{2}\,\frac{v_0}{\vo}\left[ \erf\left(z\right) + \erf\left(x_{-}\right) \right]
			-\left(1 +x_{-}\right)\,e^{-z^2}
			& \text{for }\vesc - \vo < v_q < \vesc + \vo\\
		0	& \text{for }\vesc + \vo < v_q
		\end{cases}\,,
        \mkern-60mu
    \end{aligned}
    \end{align}
where $x_{+}\equiv(\vo + v_q)/v_0$, $x_{-}\equiv(\vo - v_q)/v_0$, $z\equiv\vesc/\vo$, and the normalization constant $\N$ is provided in \cref{app:vDM}.

\subsection{Velocity-dependent \texorpdfstring{$B$}{B}}
For the models studied in this work, the DM-electron scattering amplitude $\mathscr{M}$ is at most linear in $\vec v$. Consequently, the function $B$ can be at most quadratic in $\vec v$:
	\begin{align}
    \label{eq:Bv}
	B(\omega, q, v) = B^{(0)}(\omega, q) + B^{(2)}(\omega, q)\,v^2\,.
	\end{align}
The absence of terms linear in $\vec v$ follows from the fact that to contribute to a scalar function $B$, they must be contracted with $\vec q$. This product can be then expressed in terms of $\omega$ and $q$, see \cref{eq:omegaApp}.
The contribution to $\Gamma$ from $B^{(0)}$ has been discussed in the previous subsection. In general,	
    \begin{align}
    \begin{aligned}
	\Gamma
	&= \int {\mathrm d}\vec q\,{\mathrm d}\vec v\,f(\vec v)\,B(\omega, q, v)\\
	&= \int {\mathrm d}\vec q\,{\mathrm d}\vec v\,f(\vec v)\,[B^{(0)}(\omega,q) + B^{(2)}(\omega,q)\,v^2]\\
	&= 4\pi\int_0^\infty q^2{\mathrm d}q\int_0^\infty {\mathrm d}\omega\,
		\left[\rho^{(0)}(\omega,q)\,B^{(0)}(\omega,q) +\rho^{(2)}(\omega,q)\,B^{(2)}(\omega,q)\right]\,,
	\end{aligned}
	\end{align}
where $\rho^{(2)}$ is defined analogously to $\rho^{(0)}$, but with an additional $v^2$ factor:
	\begin{align}
	\rho^{(2)}(\omega,q)
	= \frac{\pi}{q}\int_{v_q}^\infty v^3\,{\mathrm d}v\int_{-1}^1 {\mathrm d}\cos\alpha\,f(\vec v)\,.
	\end{align}
By performing the above integral analytically, we obtain
	\begin{align}
    \begin{aligned}
	\rho^{(2)}(\omega,q)
	&= \frac{\N\pi}{q}\int_{v_q}^\infty v^3\,{\mathrm d}v\int_{-1}^1 {\mathrm d}\cos\alpha\,\exp\left[-\frac{(\vv + \vvo)^2}{v_0^2}\right]\,\theta(\vesc-|\vv+\vvo|)\\[5pt]
	&=	\frac{\N \pi v_0^4}{4q}
		\begin{cases}
		\frac{\sqrt\pi}{2}\frac{v_0}{\vo}\left(1+2\frac{\vo^2}{v_0^2}\right)
			\left[
				\erf\left( x_{+} \right)
				+ \erf\left( x_{-} \right)
			\right]
			&\multirow{4}{*}{for $
			v_q < \vesc - \vo$}\\
			\quad
			+ \left(1 - \frac{v_q}{\vo}\right)e^{-x^2_{+}}
			+ \left(1 + \frac{v_q}{\vo}\right)e^{-x^2_{-}}\\
			\quad
			- 4\left(1 + \frac{\vesc^2}{v_0^2} + \frac13 \frac{\vo^2}{v_0^2}\right)e^{-\frac{\vesc^2}{v_0^2}}\\[10pt]
		\frac{\sqrt\pi}{2}\frac{v_0}{\vo}\left(1+2\frac{\vo^2}{v_0^2}\right)
			\left[
				\erf\left( \frac{\vesc}{v_0} \right)
				+ \erf\left(x_{-} \right)
			\right]
			&\multirow{3}{*}{for $\vesc - \vo < v_q < \vesc + \vo$}\\
			\quad
			+ \left(1 + \frac{v_q}{\vo}\right)e^{-x^2_{-}}\\
			\quad
			- 2\left[
				1 + \frac12\,z
				+ \frac13 \frac{ (\vesc + \vo)^3 - v_q^3}{v_0^2\,\vo}
			\right]
			e^{-\frac{\vesc^2}{v_0^2}}\\[10pt]
		0	& \text{for }\vesc + \vo < v_q
		\end{cases}
    \end{aligned}\,,
    \mkern-100mu
	\end{align}
where, again, $x_{+}\equiv(\vo + v_q)/v_0$, $x_{-}\equiv(\vo - v_q)/v_0$, $z\equiv\vesc/\vo$, and $\N$ is defined in \cref{app:vDM}.

\section{\texorpdfstring{$F$}{F}-functions}
\label{app:F}

Below, we list explicit expressions for the functions $F_{n_0}^{ss'}$, $F_{n_A}^{ss'}$, $\vec F_{j_5}^{ss'}$, $\vec F_{j_M}^{ss'}$, and $\vec F_{j_E}^{ss'}$. They are given by
\begin{subequations}
\label{eq:F}
\begin{align}
F_{n_0}^{ss'} = &~\xi_\chi^{s'\dagger}\left[ c_1 + i \left( \frac{\vec q}{m_e}  \times \vec{v}_\chi^\perp \right) \cdot  \vec S_\chi  c_5
+ \vec{v}_\chi^\perp \cdot \vec S_\chi \right.  c_8 + \left. i \frac{\vec q}{m_e} \cdot \vec S_\chi c_{11} \right] \xi^{s}\,, \\
F_{n_A}^{ss'} =&~-\frac{1}{2} \xi_\chi^{s'\dagger} \left[ c_7   +i \frac{\vec q}{m_e} \cdot \vec S_\chi~ c_{14} \right] \xi_\chi^s\,, \\
\vec F_{j_5}^{ss'} =
&~\frac{1}{2}\xi_\chi^{s'\dagger} \left[\phantom{+}
i \frac{\vec q}{m_e} \times \vec{v}_\chi^\perp~ c_3
+ \vec S_\chi c_4
+ \frac{\vec q}{m_e}~\frac{\vec q}{m_e} \cdot \vec S_\chi c_6  \right.\notag\\
&\mkern60mu+ \vec{v}_\chi^\perp c_7 + i \frac{\vec q}{m_e} \times \vec S_\chi c_9 + i \frac{\vec q}{m_e}c_{10} \\
&\mkern60mu+  \left. 
 \vec{v}_\chi^\perp \times \vec S_\chi c_{12}  + i \vec{v}_\chi^\perp \frac{\vec q}{m_e} \cdot \vec S_\chi ~ c_{14}  
+\frac{\vec q}{m_e} \times \vec{v}_\chi^\perp~ \frac{\vec q}{m_e} \cdot \vec S_\chi ~ c_{15}  \right] \xi_\chi^s\,,\notag\\
\vec F_{j_M}^{ss'} =&~\xi_\chi^{s'\dagger} \left[  i \frac{\vec q}{m_e}  \times \vec S_\chi c_5 - \vec S_\chi c_8 \right] \xi_\chi^s \,, \\
\vec F_{j_E}^{ss'} =&~\frac{1}{2} \xi_\chi^{s'\dagger}\left[  \frac{\vec q}{m_e} ~ c_3 +i \vec S_\chi c_{12} 
-i \frac{\vec{q}}{m_e} \frac{\vec q}{m_e} \cdot \vec S_\chi  c_{15} \right] \xi_\chi^s\,,
\end{align}
\end{subequations}
where
\begin{align}
\vec{v}_\chi^\perp = \left(\frac{\vec p + \vec p'}{2 m_\chi}\right) = \vec v - \frac{\vec q}{2 m_\chi}\,,
\end{align}
$\vec v=\vec p/m_\chi$, $\vec q=\vec p-\vec p'$ is the momentum transferred to the electron and ${\vec S_{\chi}\equiv\vec\sigma_{\chi}/2}$ ${(\vec S_e\equiv\vec\sigma_e/2)}$ denotes the spin matrix corresponding to the dark particle (electron).
Elements of $\mathcal{F}_{ab}$ are defined as
\begin{align}
\mathcal{F}_{ab}
\equiv \frac12\sum_{ss'} F_a^{ss'*}F_b^{ss'}\,.
\end{align}
The elements relevant for the anapole, electric dipole and magnetic dipole models investigated in this work can be explicitly expressed as
\begin{subequations}
\label{eq:F2App}
\begin{align}
\mathcal{F}_{n_0n_0}		&= c_1^2 + \frac14\left|\frac{\vec q}{m_e}\times\vvx\right|^2 c_5^2 + \frac14 \vvxSq c_8^2 + \frac14 \frac{q^2}{m_e^2} c_{11}^2\,,\\
\sum_k\mathcal{F}_{j_{5_k}j_{5_k}}
    &= \frac14\left[\phantom{+}
        \left|\frac{\vec q}{m_e}\times\vvx\right|^2 c_3^2 + \frac34 c_4^2 + \frac{q^4}{4\,m_e^4} c_6^2 + \vvxSq c_7^2 +\frac{q^2}{2m_e^2} c_9^2
        \right.\notag\\
    &\mkern50mu
        + \frac{q^2}{m_e^2} c_{10}^2 + \frac{\vvxSq}{2} c_{12}^2 + \frac{q^2}{4m_e^2}\vvxSq c_{14}^2 + \left|\frac{\vec q}{m_e}\times\vvx\right|^2 \frac{q^2}{4m_e^2} c_{15}^2\\
    &\mkern50mu\left.
        +\, \frac{q^2}{2m_e^2}c_4c_6
        - \frac12\left|\frac{\vec q}{m_e}\times\vvx\right|^2 c_{12}c_{15}
        \right]\,,\notag\\
\mathcal{F}_{j_{M_k}j_{M_{\ell}}}	&= \frac{q^2\delta_{k\ell} - q_kq_{\ell}}{4m_e^2}c_5^2 + \frac14 c_8^2\delta_{k\ell} - \frac{i}{2}\epsilon_{k\ell m}\frac{q_m}{m_e} c_5 c_8\,,\\
\mathcal{F}_{j_{M_k}n_0}	&= -\frac14\vx{k}c_8^2 - \frac{i}{2} \left(\frac{\vec q}{m_e}\times\vvx\right)_k c_5c_8 - \frac{i}{4}\frac{q_k}{m_e}c_8c_{11}\,.
\end{align}
\end{subequations}
Note that the expressions $(\vvx)^2$ and $\left|\frac{\vec q}{m_e}\times\vvx\right|^2$ can be expressed in terms of $v\equiv |\vec v|$, $q\equiv |\vec q|$ and $v_q\equiv \vec v\cdot\frac{\vec q}{q}$ as
\begin{align}
(\vec v_\chi^\perp)^2 &= v^2 - v_q\,\frac{q}{m_\chi} + \frac{q^2}{4\,m_\chi^2}\,,&
\left|\frac{\vec q}{m_e}\times\vvx\right|^2 &= \frac{q^2}{m_e^2}\,(v^2 - v_q^2)\,.
\end{align}
For completeness, we also list the other elements:
\begin{subequations}
\label{eq:F2AppCont}
\begin{align}
\mathcal{F}_{n_An_A}		&= \frac14\left(c_7^2 + \frac14\frac{q^2}{m_e^2} c_{14}^2\right)\,,\\
\mathcal{F}_{j_{E_k}j_{E_{\ell}}}	&= \frac14\left(
    \frac{q_kq_{\ell}}{m_e^2}c_3^2 + \frac14\delta_{k\ell}c_{12}^2 + \frac{q^2}{4m_e^2}\frac{q_kq_{\ell}}{m_e^2}c_{15}^2
    - \frac{q_kq_{\ell}}{2m_e^2} c_{12}c_{15}
    \right)\,,\\
\mathcal{F}_{j_{E_k}n_A}	&= \frac14\left[ -\frac{q_k}{m_e}c_3c_7 - \frac{q_k}{4m_e}c_{12}c_{14} + \frac{q_k}{4m_e}\frac{q^2}{m_e^2}c_{14}c_{15} \right]\,,\\
\mathcal{F}_{n_Aj_{5_k}}	&= -\frac14\vx{k}c_7^2 - \frac{q^2}{16m_e^2}\vx{k}c_{14}^2 - \frac{i}{4} \left(\frac{\vec q}{m_e}\times\vvx\right)_k c_3c_7
        - \frac{i}{4}\frac{q_k}{m_e}c_7c_{10}\\
    &\mkern40mu
        + \frac{i}{16}\frac{q_k}{m_e}c_4c_{14}
        + \frac{i}{16}\frac{q^2}{m_e^2}\frac{q_k}{m_e}c_6c_{14}
        + \frac{i}{16}\frac{q^2}{m_e^2}\left(\frac{\vec q}{m_e}\times\vvx\right)_kc_{14}c_{15}\,,\notag\\
\epsilon_{k\ell m}\mathcal{F}_{j_{5_{\ell}}j_{E_m}}
    &= \frac14\left[
        i\,\frac{q_k\vq\cdot\vvx - q^2\vx{k}}{m_e^2}\,c_3^2 - \frac{i}{2}\vx{k}c_{12}^2
        + \frac{i}{4}\,\frac{\vq\cdot\vvx q_k - q^2\vx{k}}{m_e^2}\,\frac{q^2}{m_e^2}c_{15}^2\right.\notag\\
    &\mkern40mu
        - \left(\frac{\vec q}{m_e}\times\vvx\right)_k\,c_3c_7 - \frac{q_k}{2m_e}c_9c_{12}
        - \frac{5i}{4}\,\frac{\vec q\cdot\vvx q_k - q^2\vx{k}}{m_e^2}\,c_{12}c_{15}\\
    &\mkern40mu\left.{}
        - \left(\frac{\vec q}{m_e}\times\vvx\right)_k\,c_{12}c_{14} + \frac{q^2}{4m_e^2}\left(\frac{\vec q}{m_e}\times\vvx\right)_k\,c_{14}c_{15}
    \vphantom{\frac{\vvx}{m_e^2}}\right]\,.\notag
\end{align}
\end{subequations}

\section{On the isotropy and \texorpdfstring{$T\to 0$}{T -> 0} assumptions}
\label{app:assumptions}
\subsection{Isotropy of the material}
\label{app:isotropy}
Following \cite{Catena:2024rym}, in our derivation we assume isotropy of the detector material. This assumption allows us to:
	\begin{itemize}
    \item integrate out most of the angular variables in \cref{app:intdv},
	\item neglect the screening corrections to the transverse responses \cite{Catena:2024rym}.
	\end{itemize}
Although the assumption of isotropy is not always strictly satisfied, it is often satisfied approximately, to the extent that allows us to neglect the local-field \cite{Knapen:2021run} and the screening \cite{Catena:2021qsr} corrections.

For isotropic and non-spin-polarized materials, the results provided in this work are exact. As we point out above \cref{eq:epsilon}, below \cref{eq:Fjn}, and above \cref{eq:appendix:gamma}, the assumption of isotropy is mathematically equivalent to averaging over detector's orientation. Thus, for non-isotropic materials, the results correspond to an expected value given the detector's orientation is chosen randomly, or to a situation in which the orientation changes cyclically (e.g., the detector is rotating). This standard approach has been also used and extensively discussed in, e.g., \cite{Lasenby:2021wsc,Essig:2015cda,Catena:2021qsr}.

\subsection{Temperature dependence of the results}
\label{app:temperature}
In our results, the thermal corrections are encoded in the exponential term $e^{-\beta\omega}$ in \cref{eq:chi1}. The detectors of interest are assumed to operate in temperatures not exceeding the room temperature, equivalent to ca.~$0.025~\text{eV}$. This value is very small in comparison to the smallest energies considered in our manuscript, e.g., the one-electron ionization threshold of germanium is $0.67~\text{eV}$. Hence, the exponential term in \cref{eq:chi1} is completely negligible in comparison with 1, which justifies using the limit $T\to 0$.
   
	For metals, not considered in our manuscript, those effects could be of some relevance because of the absence of the band gap whose energy would have to be exceeded to induce a signal. For instance, for energy transfer equal to $1~\text{meV}$ , the value of the $(1-e^{-\beta\omega})$ term in \cref{eq:chi1} becomes $0.04\ll 1$. Nevertheless, to obtain the total interaction rate we integrate over the whole range of allowed $\omega$'s, so that the higher values, with negligible thermal term, should dominate the result and the thermal correction should not significantly affect the ratio between the actual interaction rate and the theoretical optimum.
	
	One should definitely take into account the thermal effects described by the $e^{-\beta\omega}$ term when considering phononic detection channel, for which even the smallest energies may be relevant and could provide a measurable signal. This case is, however, beyond the scope of this work.

\bibliographystyle{JHEP}
\bibliography{ref,ref2}

\end{document}